\documentclass[12pt,thmsa]{article}
\usepackage{sw20lart}

\input tcilatex
\QQQ{Language}{
American English
}

\begin{document}

\title{Linear Relaxation Processes Governed by Fractional Symmetric Kinetic
Equations}
\author{A. V. Chechkin and V. Yu. Gonchar \\
%EndAName
Kharkov Institute for Theoretical Physics\\
National Science Center\\
``Kharkov Institute of Physics and Technology'',\\
Akademicheskaya st.1, Kharkov 310108 Ukraine\\
and \\
Institute for Single Crystals\\
National Acad. Sci. Ukraine,\\
Lenin ave. 60, Kharkov 310001, Ukraine}
\maketitle

\begin{abstract}
We get fractional symmetric Fokker - Planck and Einstein - Smoluchowski
kinetic equations, which describe evolution of the systems influenced by
stochastic forces distributed with stable probability laws. These equations
generalize known kinetic equations of the Brownian motion theory and contain
symmetric fractional derivatives over velocity and space, respectively. With
the help of these equations we study analytically the processes of linear
relaxation in a force - free case and for linear oscillator. For a weakly
damped oscillator we also get kinetic equation for the distribution in slow
variables. Linear relaxation processes are also studied numerically by
solving corresponding Langevin equations with the source which is a discrete
- time approximation to a white Levy noise. Numerical and analytical results
agree quantitatively.

PACS: 05.10 Gg, 05.40. Fb
\end{abstract}

\newpage

\newpage

\section{Introduction}

Studies of evolution of various systems under the influence of external
stochastic forces constitute content of a large section of statistical
physics. It has a great variety of applications in physics, chemistry,
biology, economy and sociology, see, e.g., \cite{Kampen}, \cite
{Klimontovich1}, \cite{Klimontovich2}. The most known problem is the
description of Brownian motion. In classical formulation the Brownian motion
manifests itself as unceasing chaotic motion of small macroscopic particles
in a liquid or a gas. The Brownian motion is explained by the presence of
unbalanced ``pushes'' of surrounding atoms and, hence, reveals atomic
structure of continuous medium, in which the motion occurs.

The achievements of probability theory serve as a mathematical basis of the
kinetic theory of the Brownian motion. They are as follows:

(i) the Central Limit Theorem, which justifies Gaussianity of stochastic
force acting on Brownian particle;

(ii) the theory of Markovian stochastic processes; an important assumption
used at the kinetic description of the Brownian motion is that the behavior
of particle at a given instant depends only on instantaneous values of
physical parameters, but not on their previous history;

(iii) studies of stochastic Gaussian processes, first of all, the work of
Bachelier (1900), who was the first to study a continuous stochastic
Gaussian process with independent increments, and the work of Wiener (1927),
who has given a rigorous mathematical formulation of this process and has
studied properties of its sample paths.

The basic equations of the kinetic theory of Brownian motion are the Fokker
- Planck equation for the probability density function (PDF) $f(x,v,t)$ in
the phase space of coordinates and velocities, and the Einstein -
Smoluchowski equation for the PDF $f(x,t)$ in the real space. The relaxation
in the phase space can occur in two steps: the first one, ``fast'' stage, at
which the relaxation over velocities occurs, and Maxwellian PDF is
established, and the second one, ``slow'' diffusion stage, at which
relaxation in the real (coordinate) space occurs. If in the system
considered the physical conditions are such that the two stages of
relaxation can be separated, then it is possible to pass from the Fokker -
Planck equation to the Einstein - Smoluchowski equation and describe the
system on the diffusion stage with a simpler equation. In more detail the
transition to the diffusion stage is discussed in the article by
Chandrasekhar \cite{Chandrasekhar}, which contains a brilliant presentation
of classical theory of the Brownian motion, as well as in the monograph \cite
{Klimontovich2} containing a modern presentation of the theory of the
Brownian motion with inclusion of the motion in nonlinear open systems.

Beginning from the second half of 80-th the term ``Levy motion'' is becoming
widely used in statistical physics, in particular, in problems of anomalous
diffusion, when characteristic displacement of diffusive particle grows as $%
t^{\mu }$ , $\mu \neq 1/2$ (the case $\mu =1/2$ corresponds to the classical
Brownian motion). Levy anomalous diffusion appears in different areas of
physics, namely, turbulence \cite{Klafter}, solid and amorphous states
physics \cite{Bouchaud}, chaotic dynamics \cite{Zaslavsky}, plasma physics 
\cite{Zimbardo} etc. It is worthwhile to mention also non - physical areas,
e. g., biology and physiology \cite{West}, and the theory of finance \cite
{Bouchaud2}. The Levy motion can be considered as a generalization of the
Brownian motion. Indeed, the mathematical foundation of the generalization
are remarkable properties of stable probability laws, whose theory has begun
from the works of Levy and Khintchine \cite{Levy}. From the limit theorems
point of view the stable probability laws are the generalization of widely
used Gaussian law. Namely, stable laws are the limit ones for the
probability laws of (properly normalized) sums of independent identically
distributed (i.i.d.) random variables \cite{Gnedenko}. Therefore, these laws
(as the Gaussian one) occur, when evolution of a physical system or the
result of an experiment are determined by the sum of a large number of
identical independent random factors. An important distinction of stable
PDFs is the power law tail decreasing as $\left| x\right| ^{-1-\alpha
},\,\;x\rightarrow \infty ,$ $\alpha $ is the Levy index, $0$ $<$ $\alpha <$ 
$2$. Hence, the PDF moments of the order $q\geq \alpha $ diverge.

The above mentioned properties of stable PDFs allow one to get a simple
intuitive basis for anomalous diffusion in the framework of the model of
independent random ``jumps'' \cite{Bouchaud}. However, in order to construct
a consistent theory of the Levy motion it is necessary to get the kinetic
equations, which generalize those of the Brownian motion, namely, the
Fokker-Planck and the Einstein-Smoluchowski equations. It is clear also from
the very simple arguments that these equations must contain space and/or
time fractional derivatives. During last two decades a few monographs solely
devoted to the theory of fractional calculus were appeared, see, e. g., \cite
{Samko}, and an extensive treatment of fractional order differential
equations applied to heat and mass transfer were given \cite{Babenko}.
Different forms of diffusion - like equations with fractional derivatives
were proposed \cite{Wyss}, \cite{Schneider1}, \cite{Nonnenmacher}, \cite
{Jumarie}, \cite{Saichev}, \cite{Mainardi2}. These equations were applied,
in particular, for the description of anomalous diffusion on random fractals 
\cite{Giona}, \cite{Roman}, and in chaotic Hamiltonian systems, for which
the orders of space and time fractional derivatives are determined by
delicate properties of the phase space \cite{Zaslavsky}, \cite{Zaslavsky2}.
We also refer to the paper \cite{Mainardi1}, where the general description
of the fractional relaxation - oscillation and fractional diffusion - wave
phenomena was provided with adoption a mathematical approach to the
fractional calculus that is as simple as possible.

Our paper deals with fractional generalizations of the classical
Fokker-Planck and Einstein-Smoluchowski equations describing relaxation in
the phase and real spaces, respectively. We follow classical approach \cite
{Chandrasekhar}, which is used to derive kinetic equations for the Brownian
motion as well as an approach used in Ref. \cite{Peseckis}, where, as far as
we know, the fractional kinetic equation for the phase space PDF was
proposed for the first time. Through the paper we restrict ourselves by a
one - dimensional coordinate space and two - dimensional (coordinate plus
velocity) phase space. The generalization on two - and three - dimensional
real space cases are discussed at the end of the article. Besides, we
restrict ourselves by symmetric fractional kinetic equations, that is,
those, which include symmetric fractional derivatives (see below).

Thus, in our paper we, at first, derive kinetic equations with symmetric
fractional derivatives, which generalize Fokker - Planck and Einstein -
Smoluchowski equations to the case of the Levy motion. These equations are
called fractional symmetric Fokker - Planck equation (FSFPE) and fractional
symmetric Einstein - Smoluchowski equation (FSESE), respectively. Secondly,
we use these equations for studying classical problems of linear relaxation,
namely, relaxation in a force - free case and relaxation of a linear Levy
oscillator.

It is worthwhile to note that the force - free relaxation in a spatially
homogeneous case as well as relaxation of a linear oscillator were first
studied in Ref. \cite{West2}. In this paper the equations for characteristic
functions were obtained by solving corresponding Langevin equations with
subsequent averaging the Liouville equation for the phase space density.
Besides, with the help of fractional kinetic equation for the diffusion
stage of relaxation the linear oscillator was considered in Ref. \cite
{Jespersen}. With the help of FSFPE and FSESE we study in detail both
``fast'' and ``slow'' stages of a linear relaxation and demonstrate a
transition from the first level of description (the use of FSFPE) to the
second one (the use of FSESE).

Further, we point attention on a description of two limit cases for the
oscillator, namely, overdamped oscillator and a weakly damped oscillator.
Both cases are very important when studying the Levy motion in nonlinear
open systems. We propose a new kinetic equation for the weakly damped linear
oscillator and study its solutions. We also solve numerically the Langevin
equations, which correspond to fractional kinetic equations. We demonstrate
numerical results for the linear relaxation problems which are solved
analytically, and show a good agreement between the results of kinetic
theory and the results of numerical modelling.

The paper is organized as follows. In Sec. 2 we derive fractional
generalizations of the Fokker - Planck and the Einstein - Smoluchowski
equations following the approaches of \cite{Chandrasekhar}, \cite{Peseckis}.
In Sec. 3 we investigate relaxation in a real space for the force - free
Levy motion and for the Levy linear oscillator. In Sec. 4 we investigate
both relaxation problems in the phase space. In Sec. 5 we consider
relaxation for the limit cases of the Levy oscillator, namely, overdamped
and weakly damped oscillators. Finally, the conclusions and discussion are
presented in Sec. 6.

\section{Fractional Fokker-Planck and fractional Einstein-Smoluchowski
equations}

It is known that usually the derivation of the Fokker - Planck equation is
based on the assumption of the finiteness of the second moment of the PDF.
Since this assumption breaks down for the stable PDFs, we find that the
classical approach used by Chandrasekhar \cite{Chandrasekhar} can be adopted
for the derivation without using the finiteness of the second moment. A
similar treatment was undertaken in Ref. \cite{Fogedby}, where the
discussion of Ref. \cite{Stratonovich} was adopted for the purpose of
getting kinetic equation in coordinate space.

\subsection{Fractional Fokker - Planck equation}

Like in the Brownian motion theory, the initial equations in our approach
are as follows:

1) integral equation for the PDF $f(x,v,t)$ of the Markovian stochastic
process in the phase space: 
\begin{eqnarray}
f(x,v,t+\Delta t) &=&\int \int d(\Delta x)d(\Delta v)f(x-\Delta x,v-\Delta
v,t)\times  \nonumber \\
&&\times \Psi (x-\Delta x,v-\Delta v;\Delta x,\Delta v,\Delta t)\,, 
\tag{2.1}
\end{eqnarray}
where $\Psi (x,v;\Delta x,\Delta v,\Delta t)$ $\,$is the transition
probability, that is, the probability for coordinate $x$ to get increment $%
\Delta x,$and for the velocity $v$ the increment $\Delta v$ during the
interval $\Delta t,$ as well;

2) the Langevin equations 
\begin{eqnarray}
\frac{dx}{dt} &=&v\quad ,  \nonumber \\
\frac{dv}{dt} &=&-\nu v+F+A(t)\quad ,  \tag{2.2}
\end{eqnarray}
where $\nu $ is the friction coefficient, which is assumed to be independent
of $v$, $F$ is the regular external force, and $A(t)$ is the fluctuation
component of the external force.

Further in the theory of Brownian motion, by following traditional
assumptions \cite{Chandrasekhar}, one gets expressions for the increments of
coordinate and velocity during time interval $\Delta t$, which is longer
than characteristic time intervals of $A(t)$, though being shorter than time
intervals during which physical parameters change appreciably: 
\begin{eqnarray}
\Delta x &=&v\Delta t\quad ,  \nonumber \\
\Delta v &=&-(\nu v-F)\Delta t+B(\Delta t)\quad .  \tag{2.3}
\end{eqnarray}
Here $B(\Delta t)=\int\limits_{t}^{t+\Delta t}A(t^{^{\prime }})dt^{^{\prime
}}$ is a non-stationary stochastic process, which is assumed to be
homogeneous Gaussian process with independent increments possessing PDF 
\begin{equation}
w(B(\Delta t))=\frac{1}{\sqrt{4\pi D\Delta t}}\exp \left( -\frac{(B(\Delta
t))^{2}}{4D\Delta t}\right) \quad .  \tag{2.4}
\end{equation}
The Central Limit Theorem serves as a mathematical justification of this
assumption. In accordance with stated above, we make a generalization of
Chandrasekhar's approach with the help of generalization of the Central
Limit Theorem to the case of i.i.d. variables having infinite variances.
Indeed, Levy and Khintchine \cite{Levy} discovered a class of stable
probability laws. They are limit ones for the probability laws of normalized
sums of i.i.d. random variables. Each stable law with a characteristic Levy
index (0 $<\alpha $ $<$ 2) possesses finite moments of orders $q$, 0 $<$ $q$ 
$<\alpha $ , but infinite moments for higher orders. The Gaussian law is
also stable one with a characteristic index $\alpha $= 2, and it possesses
moments of all orders. Returning to Eq. (2.4) we note that in the theory of
stochastic processes the corresponding generalization of homogeneous
Gaussian process $B(t)$ with independent increments are the Levy stable
processes $L(t)$, having the characteristic function (we restrict ourselves
by symmetric stable laws) \cite{Skorokhod} 
\begin{equation}
\hat{w}_{L}(k,\Delta t))=\left\langle e^{ikL}\right\rangle =\exp (-D\left|
k\right| ^{\alpha }\Delta t)\quad ,  \tag{2.5}
\end{equation}
where D $>$ 0, $(D\Delta t)^{1/\alpha }$ is called the scale parameter. At $%
\alpha $= 2 one has Gaussian process $B(t)$. Above statements justifies
expediency of generalization $B(\Delta t)\rightarrow L(\Delta t)$ in Eqs.
(2.3).

With Eqs. (2.3) and (2.5) the transition probability, see Eq. (2.1) is 
\begin{equation}
\Psi (x,v;\Delta x,\Delta v)=\psi (x,v;\Delta v)\delta (\Delta x-v\Delta
t)\quad ,  \tag{2.6}
\end{equation}
where $\delta $ implies the Dirac delta-function, and 
\begin{equation}
\psi (x,v,\Delta v)=\int\limits_{-\infty }^\infty \frac{dk}{2\pi }\exp
\left[ -ik(\Delta v+\nu v\Delta t-F\Delta t)-D\left| k\right| ^\alpha \Delta
t\right]  \tag{2.7}
\end{equation}
is the transition probability in the velocity space.

We insert Eqs. (2.6) and (2.7) into Eq. (2.1), expand the left and the right
sides of the equation obtained into series of $\Delta t$ and set $\Delta
t\rightarrow 0$. As the result we get 
\[
\frac{\partial f}{\partial t}+v\frac{\partial f}{\partial x}= 
\]
\begin{eqnarray}
&=&-\int d(\Delta v)f(x,v-\Delta v,t)\int\limits_{-\infty }^\infty \frac{dk}{%
2\pi }\exp (-ik\Delta v)\times  \nonumber \\
&&\quad \times \left[ -ikF+ikv(v-\Delta v)+D\left| k\right| ^\alpha \right]
\quad .  \tag{2.8}
\end{eqnarray}
Now we turn to the physical space by making an inverse Fourier transform
over velocity in the right - hand side of Eq. (2.8). We treat each term in
the square brackets separately. The first and the second terms, being
``classical '' ones, are transformed trivially, leading to 
\[
-F(x,t)\frac{\partial f}{\partial v} 
\]

and 
\[
\nu \frac \partial {\partial v}(vf)\qquad , 
\]
respectively. We turn to the last term, which can be written as 
\begin{equation}
-D\int_{-\infty }^\infty \frac{d\kappa }{2\pi }e^{-i\kappa x}\int_{-\infty
}^\infty \frac{dk}{2\pi }e^{-ikv}\left| k\right| ^\alpha \widehat{f}(\kappa
,k,t)\qquad ,  \tag{2.9}
\end{equation}
where $\stackrel{\wedge }{f}(\kappa ,k,t)$ is the characteristic function, 
\begin{equation}
f(x,v,t)=\int_{-\infty }^\infty \frac{d\kappa }{2\pi }e^{-i\kappa
x}\int_{-\infty }^\infty \frac{dk}{2\pi }e^{-ikv}\widehat{f}(\kappa
,k,t)\quad .  \tag{2.10}
\end{equation}

We use the symmetric fractional derivative of any order $\alpha >0$, which
may be defined, for a ``sufficiently well - behaved'' function $\phi
(v),v\in \mathbf{R,}$ as the (pseudo - differential) operator characterized
in its Fourier representation by 
\begin{equation}
\frac{d^\alpha }{d\left| v\right| ^\alpha }\phi (v)\div -\left| k\right|
^\alpha \widehat{\phi }(k),\quad v,k\in \mathbf{R},\alpha >0.  \tag{2.11}
\end{equation}
Here at the left - hand side we adopt the notation introduced in Ref. \cite
{Saichev}.

In order to treat this kind of fractional derivative properly, we remind the
definition of the left - and right - side Liouville derivatives on the
infinite axis \cite{Samko}, 
\begin{equation}
\mathcal{D}_{+}^\alpha \phi (v)=\frac 1{\Gamma (1-\alpha )}\frac
d{dv}\int_{-\infty }^v\frac{\phi (\xi )d\xi }{(x-\xi )^\alpha }\quad , 
\tag{2.12}
\end{equation}
\begin{equation}
\mathcal{D}_{-}^\alpha \phi (v)=-\frac 1{\Gamma (1-\alpha )}\frac
d{dv}\int_v^\infty \frac{\phi (\xi )d\xi }{(\xi -x)^\alpha }\quad , 
\tag{2.13}
\end{equation}
where $0<\alpha <1.$ For $\alpha \geq 1$%
\begin{equation}
\mathcal{D}_{\pm }^\alpha \phi (v)=\frac{(\pm 1)^n}{\Gamma (n-\alpha )}\frac{%
d^n}{dv^n}\int_0^\infty \xi ^{n-\alpha -1}\phi (v\mp \xi )d\xi \quad , 
\tag{2.14}
\end{equation}
$n=\left[ \alpha \right] +1$, here the square brackets denote the integer
part of the order of the derivative. The derivatives (2.12) - (2.14) are
characterized in their Fourier representation by 
\begin{equation}
\mathcal{D}_{\pm }^\alpha \phi (v)\div (\mp ik)^\alpha \widehat{\phi }%
(k),\quad \alpha \geq 0,  \tag{2.15}
\end{equation}
where 
\[
(\mp ik)^\alpha =\left| k\right| ^\alpha \exp \left( \mp \frac{i\alpha \pi }%
2sgnk\right) . 
\]
Thus, the symmetric space - fractional derivative (2.11) can be written as 
\begin{equation}
\frac{d^\alpha }{d\left| v\right| ^\alpha }\phi (v)=-\frac 1{2\cos (\pi
\alpha /2)}\left[ \mathcal{D}_{+}^\alpha \phi (v)+\mathcal{D}_{-}^\alpha
\phi (v)\right] \quad ,  \tag{2.16}
\end{equation}
where $\alpha \neq 1,3,...$

Now we are able to return to Eq. (2.8) and write the kinetic equation for
the PDF $f(x,v,t)$ in the phase space as

\begin{equation}
\frac{\partial f}{\partial t}+v\frac{\partial f}{\partial x}+F\frac{\partial
f}{\partial v}=\nu \frac \partial {\partial v}(vf)+D\frac{\partial ^\alpha }{%
\partial \left| v\right| ^\alpha }f  \tag{2.17}
\end{equation}
where the last term is defined through Eqs. (2.11) - (2.16). This is a
fractional Fokker - Planck equation for the Levy motion. At $\alpha =2$ this
is an ordinary Fokker - Planck equation for the Brownian motion.

Notice that Eq. (2.17) becomes meaningless when $\alpha $ is an odd integer
number. That is why, for our range of interest $0<\alpha \leq 2$, the
particular case $\alpha =1$ must be treated separately. However, if one uses
the Fourier transform over velocity when solving a particular problem, this
case is not singled out.

\subsection{Fractional Einstein - Smoluchowski equation}

At the Brownian motion description, along with the relaxation parameter $%
1/\nu $, another relaxation parameter exists, which characterizes diffusion
in a real space. If characteristic time of this process is much greater than 
$1/\nu $, then it is possible to pass from the Fokker - Planck equation for
the PDF $f(x,v,t)$ to the Einstein - Smoluchowski equation for a simpler PDF 
$f(x,t)$.

As in previous sub - section, at the derivation of the fractional symmetric
Einstein - Smoluchowski equation we follow the reasonings used in the theory
of Brownian motion.

Instead of Eq. (2.1), an integral equation in a coordinate space serves as
an initial one,

\begin{equation}
f(x,t+\Delta t)=\int d(\Delta x)f(x-\Delta x,t)\psi (x-\Delta x;\Delta
x,\Delta t)\quad ,  \tag{2.18}
\end{equation}
where $\psi (x;\Delta x,\Delta t)$ is the transition probability, that is,
the probability for coordinate x to get increment $\Delta x$ during interval 
$\Delta t$.

In the kinetic theory of the Brownian motion the transition to the Einstein
- Smoluchowski equation corresponds to neglecting the term $dv/dt$ in the
Langevin equation (2.2) \cite{Klimontovich1}. Thus, instead of two equations
(2.2) we have a single Langevin equation, 
\begin{equation}
\frac{dx}{dt}=\frac F\nu +\frac 1\nu A(t)\quad ,  \tag{2.19}
\end{equation}
and, instead of Eqs. (2.3) we get 
\begin{equation}
\Delta x=\frac{F\Delta t}\nu +\frac 1\nu L(\Delta t)\quad ,  \tag{2.20}
\end{equation}
where $L(t)$ is a stable process with symmetric PDF and characteristic
function (2.5), as before. The transition probability follows from Eq.
(2.20), 
\begin{equation}
\psi (x,\Delta x,\Delta t)=\int\limits_{-\infty }^\infty \frac{dk}{2\pi }%
\exp \left[ -ik\left( \Delta x-\frac{F\Delta t}\nu \right) -\frac D{\nu
^\alpha }\left| k\right| ^\alpha \Delta t\right] \quad .  \tag{2.21}
\end{equation}
We insert Eq. (2.21) into Eq. (2.18), expand left - and right - hand sides
of equation obtained in Taylor series in $t$ and, then take the limit $%
t\rightarrow 0$. As the result we get fractional symmetric
Einstein-Smoluchowsky equation, 
\begin{equation}
\frac{\partial f}{\partial t}=-\frac \partial {\partial x}\left( \frac F\nu
f\right) +\frac D{\nu ^\alpha }\frac{\partial ^\alpha }{\partial \left|
x\right| ^\alpha }f\quad .  \tag{2.22}
\end{equation}
In the next Sections we give examples of relaxation processes governed by
Eqs. (2.22) and (2.17).

\section{Solutions to fractional symmetric Einstein-Smo-luchowski equation}

In this Section we consider two simple examples of relaxation processes
governed by FSESE, namely, relaxation in a force - free case and relaxation
of Levy linear oscillator.

\subsection{Force - free relaxation.}

Setting $F=0$ in Eq. (2.22), we seek for the solution of equation

\begin{equation}
\frac{\partial f}{\partial t}=\frac D{\nu ^\alpha }\frac{\partial ^\alpha }{%
\partial \left| x\right| ^\alpha }f,  \tag{3.1}
\end{equation}
with an initial condition

\begin{equation}
f\left( x,t=0\right) =\delta \left( x-x_{0}\right) .  \tag{3.2}
\end{equation}
We pass from Eqs. (3.1) and (3.2) to the equation for the characteristic
function $\widehat{f}(\kappa ,t),$

\begin{equation}
\frac{\partial \widehat{f}}{\partial t}=-\frac D{\nu ^\alpha }\left| \kappa
\right| ^\alpha \widehat{f},  \tag{3.3}
\end{equation}
with an initial condition

\begin{equation}
\widehat{f}\left( \kappa ,t=0\right) =e^{i\kappa x_0}  \tag{3.4}
\end{equation}
The solution of Eqs. (3.3) and (3.4) is

\begin{equation}
\widehat{f}\left( \kappa ,t\right) =\exp \left\{ i\kappa x_0-\frac D{\nu
^\alpha }\left| \kappa \right| ^\alpha t\right\}  \tag{3.5}
\end{equation}
Thus, in the force - free case the random process $x(t)$ is a non -
stationary Levy stable process with independent increments and with the
scale parameter 
\[
\frac{(Dt)^{1/\alpha }}\nu \quad . 
\]

In the real space the Levy stable PDFs are expressed in terms of Fox' $H$
functions \cite{Fox}. Such representation of all stable PDFs was achieved in
Ref. \cite{Schneider}. Mathematical details on $H$ functions are presented
in Refs. \cite{Mathai}, \cite{Srivastava}. However, in present paper we do
not touch a real space representation for an arbitrary $\alpha .$

Since for the stable PDFs the variance and higher moments of integer order
diverge, as statistical means characterizing the properties of these
processes the moments of fractional orders can be used \cite{West2}, \cite
{Chechkin}. In order to guarantee the reality, they must be defined for the
modulus of stochastic variable. Therefore, in case of force - free
relaxation the moments of fractional orders are

\begin{eqnarray}
M_x\left( t;q,\alpha \right) &=&\left\langle \left| x-x_0\right|
^q\right\rangle =\int_{-\infty }^\infty dx\left| x\right| ^qf\left(
x,t|0\right) =  \nonumber \\
\ &=&\left\{ 
\begin{array}{c}
((Dt)^{1/\alpha }/\nu )^qC(q;\alpha ),\quad 0<q<\alpha \\ 
\infty ,\quad q\geq \alpha
\end{array}
\right\}  \tag{3.6}
\end{eqnarray}
for $0<\alpha <2,$ whereas 
\begin{equation}
M_x\left( t;q,2\right) =\frac{(Dt)^{q/2}}{\nu ^2}C(q;2)  \tag{3.7}
\end{equation}
for $\alpha =2$ and an arbitrary $q$, where 
\[
C(q;\alpha )=\int_{-\infty }^\infty dx_2\left| x_2\right| ^q\int \frac{dx_1}{%
2\pi }\exp (-ix_1x_2-\left| x_1\right| ^\alpha )\quad . 
\]
The coefficient $C(q;\alpha )$ can be evaluated with the use of generalized
function theory \cite{West2}: 
\begin{equation}
C(q;\alpha )=\frac 2{\pi q}\sin \left( \frac{\pi q}2\right) \Gamma
(1+q)\Gamma \left( 1-\frac q\alpha \right) ,\quad 0<q<\alpha \quad . 
\tag{3.8}
\end{equation}

Equations (3.6) - (3.8) have a direct physical consequence for description
of an anomalous diffusion. Indeed, for the ordinary Brownian motion the
characteristic displacement of a particle may be written through the second
moment as 
\begin{equation}
\Delta x(t)=M_x^{1/2}(t;2,2)\propto t^{1/2}.  \tag{3.9}
\end{equation}
One may note from Eqs. (3.7), (3.8), that for the normal diffusion

\begin{equation}
M_x^{1/q}(t;q,2)\propto t^{1/2}  \tag{3.10}
\end{equation}
at any $q$ and, thus any order of the moment can serve as a measure of a
normal diffusion rate:

\begin{equation}
\Delta x(t)\approx M_x^{1/q}(t;q,2)\propto t^{1/2},  \tag{3.11}
\end{equation}
if one is interested in time - dependence of the characteristic
displacement, but not in the value of the prefactor. We remind that usually
just the time - dependence, but not the prefactor, serves as an indicator of
normal or anomalous diffusion \cite{Bouchaud}. In analogy with Eq. (3.11) it
follows from Eqs. (3.6), (3.8) that the quantity $M_x^{1/q}(t;q,\alpha )$ at
0 $<\alpha $ $<$ 2 and any $q$ $<\alpha $ can serve as a measure of an
anomalous diffusion rate:

\begin{equation}
\Delta x(t)\approx M_x^{1/q}(t;q,\alpha )\propto t^{1/\alpha },\quad
0<q<\alpha <2.  \tag{3.12}
\end{equation}
Here we have the case of a fast anomalous diffusion, or superdiffusion.

In order to illustrate how $C(q;\alpha )$ influences absolute value of
characteristic displacement $\Delta x(t)$, in Fig.1 we show the $q$ - th
order root of $C(q;\alpha )$ as a function of $q$ at different values of $%
\alpha .$ It is seen that $C^{1/q}$ weakly depends on $q$ and is near to 1,
if one does not choose $q$ close to $\alpha .$ This is especially clearly
seen for $\alpha >1.$ This figure illustrates that the value of prefactor,
at sufficiently small $q$, practically does not influence the anomalous
diffusion rate.

It is expedient to give two remarks on the above.

The first one is related to applicability of the diffusion - like equation
(3.1) at the Levy indexes less than unity. Indeed, at $0<\alpha <1$ the
characteristic displacement (3.12) grows faster than in the ballistic
regime, that is, in case of a free particle motion with a finite velocity.
This property is often proclaimed as unphysical, and the conclusion follows
about inapplicability of the diffusion - like equation (3.1) at $\alpha <1,$
see, e. g., \cite{Zolotarev}. However, as a counterargument, one could
mention the law of relative diffusion (sometimes called as Richardson law)
in the field of locally isotropic turbulence, according to which the square
of displacement grows as $t^3$, that is, faster than in the ballistic regime.

The second remark is related to the divergence of the second moment $%
M_x(t;2,\alpha )$ for the hyperdiffusion. In Ref. \cite{Jespersen}, in order
to extract the scaling form for the second moment, the ``walker'' is
enclosed in an ``imaginary growing box''. This formal procedure leads to the
diffusion scaling $t^{2/\alpha }$ for the variance, which is consistent with
Eq. (3.12), and it was implemented numerically. However, it seems that more
physically relevant procedure must take into account the finite velocity of
a diffusive particle. This problem is beyond the scope of our paper. We only
mention a recent extensive discussion on this interesting and important
theme \cite{Barkai}.

The results of numerical simulation of the Langevin equation (2.19) are
shown in Figs. 2 and 3. Here and below the stochastic source $A(t)$ is
represented in numerical simulation as a discrete approximation of a ``white
Levy noise'', that is, as a stationary consequence of independent
identically distributed variables having symmetric stable PDF with the Levy
index $\alpha $ and the scale parameter equal 1. To obtain the sequence we
use the generator, which is based on the method of inversion \cite{Kendall}
along with the Gnedenko limit theorem \cite{Gnedenko}. This generator was
described in our recent papers \cite{Chechkin1}, \cite{Chechkin2} in more
detail. Time interval between the terms of the sequence is equal unity. In a
force - free problem we estimate numerically the moments $M_{x}(t;q,\alpha )$
by averaging over realizations of $x(t).$ The total number of realization is
equal 500, each of length 512.

In Fig. 2 we show $M_x(t;q,1)$ versus $t$ at different $q$ in a log - log
scale. At $q<\alpha =1$ the dependence is well fitted by straight line whose
slope allows one to define the diffusion exponent. At $q\geq \alpha $
theoretical value of the moment is infinite, and in numerical simulation the
moment strongly fluctuates, thus it is unable to get the diffusion exponent.

In Fig. 3 we show the exponent $\mu $ in the relation 
\begin{equation}
M_{x\,}(t;q,\alpha )\propto t^{\mu }  \tag{3.13}
\end{equation}
versus the Levy index $\alpha $ of the discrete approximation of a white
noise. The order $q$ of the moment is equal $0.25$, which is smaller then
the smallest value $\alpha =1$ used in numerical simulation. Theoretical
dependence $0.25/\alpha $ is shown by dotted line. The values of $\mu $
obtained in simulations are shown by black points. A good agreement between
the theory and numerical simulation is obvious.

\subsection{ Relaxation of linear Levy oscillator}

Setting $F=-\omega ^2x$ in Eq. (2.22), we seek for the solution of equation

\begin{equation}
\frac{\partial f}{\partial t}=\frac{\omega ^2}\nu \frac \partial {\partial
x}\left( xf\right) +\frac D{\nu ^\alpha }\frac{\partial ^\alpha f}{\partial
\left| x\right| ^\alpha },  \tag{3.14}
\end{equation}
with an initial condition

\begin{equation}
f\left( x,t=0\right) =\delta \left( x-x_{0}\right) .  \tag{3.15}
\end{equation}
The equation for the characteristic function is

\begin{equation}
\frac{\partial \widehat{f}}{\partial t}=-\frac{\omega ^2}\nu \kappa \frac{%
\partial \widehat{f}}{\partial \kappa }-\frac D{\nu ^\alpha }\left| \kappa
\right| ^\alpha \widehat{f},  \tag{3.16}
\end{equation}
and an initial condition is

\begin{equation}
\widehat{f}\left( \kappa ,t=0\right) =e^{i\kappa x_0}  \tag{3.17}
\end{equation}
The solution of Eqs. (3.16) and (3.17) is obtained by the method of
characteristics:

\begin{equation}
\widehat{f}\left( \kappa ,t\right) =\exp \left\{ i\kappa x_0e^{-\frac{\omega
^2}\nu t}-D_{osc}\left( t\right) \left| \kappa \right| ^\alpha \right\}
\quad ,  \tag{3.18}
\end{equation}
where

\begin{equation}
D_{osc}\left( t\right) =\frac D{\alpha \omega ^2\nu ^{\alpha -1}}\left(
1-e^{-\frac{\alpha \omega ^2}\nu t}\right) .  \tag{3.19}
\end{equation}
This result was obtained in \cite{Jespersen}.

We see from Eqs. (3.18) and (3.19) that the relaxation of oscillator,
contrary to the force - free case, is not a Levy stable process with
independent increments. It can be named as ``stable - like'' or ``Levy -
like '' process, since there exists $\left| \kappa \right| ^{\alpha }$ in
the exponent of the characteristic function, however, the scale parameter
for the oscillator, $\left( D_{osc}\left( t\right) \right) ^{1/\alpha },$
does not follow as $t^{1/\alpha },$ which is a manifestation of the Levy
stable process with independent increments, see Eq. (2.5). The process $x(t)$
behaves as a Levy stable one only asymptotically at small times, 
\begin{equation}
t<<\tau _{x}=\frac{\nu }{\alpha \omega ^{2}}\quad ,  \tag{3.20}
\end{equation}
when the exponent in Eq. (3.19) can be expanded into power series, and we
get the force - free result with accuracy up to linear in $t$ terms
inclusively. On the other hand, at $t>>\tau _{x}$ the process $x(t)$ becomes
asymptotically stationary process with the stable PDF which does not depend
on $t$ and with the Levy index $\alpha $ and the scale parameter 
\begin{equation}
D_{osc}^{1/\alpha }(t=\infty )=\left( \frac{D}{\alpha \omega ^{2}\nu
^{\alpha -1}}\right) ^{1/\alpha }\quad .  \tag{3.21}
\end{equation}

The PDF $f(x,t)$ is expressed in terms of elementary functions in two cases:

(i) $\alpha =2$ (Brownian oscillator), 
\begin{equation}
f(x,t)=\frac 1{\sqrt{4\pi D_{osc}(t)}}\exp \left\{ -\frac{\left(
x-x_0e^{-\omega ^2t/\nu }\right) ^2}{4D_{osc}(t)}\right\} \quad ,  \tag{3.22}
\end{equation}
\[
D_{osc}(t)=\frac D{2\nu \omega ^2}\left( 1-e^{-2\omega ^2t/\nu }\right) , 
\]
and

(ii) $\alpha =1$ (Cauchy oscillator), 
\begin{equation}
f(x,t)=\frac{D_{osc}(t)/\pi }{D_{osc}^{2}(t)+\left( x-x_{0}e^{-\omega
^{2}t/\nu }\right) ^{2}}\quad ,  \tag{3.23}
\end{equation}
\[
D_{osc}(t)=\frac{D}{\omega ^{2}}\left( 1-e^{-\omega ^{2}t/\nu }\right) \quad
. 
\]
Equations (3.22) and (3.23) are the Gaussian and Cauchy PDFs, respectively,
whose scale parameters does not depend as $t^{1/2},t,$ respectively. We note
also that only for the Brownian oscillator stationary solution has Boltzmann
form, 
\begin{equation}
f_{st}(x;\alpha =2)=\sqrt{\frac{\nu \omega ^{2}}{2\pi D}}\exp \left\{ -\frac{%
\nu \omega ^{2}}{2D}x^{2}\right\} \quad .  \tag{3.24}
\end{equation}
Below we return to the problems connected with stationary solutions of
fractional kinetic equations.

It is convenient to define fractional moments after subtraction of a
stochastic quantity $x$ its regular part containing initial condition, that
is, $x_0\exp (-\omega ^2t/\nu ).$ Thus, the moments are

\begin{eqnarray}
M_x\left( t;q,\alpha \right) &=&\left\langle \left| x-x_0e^{-\omega ^2t/\nu
}\right| ^q\right\rangle =\int_{-\infty }^\infty dx\left| x\right| ^qf\left(
x,t|0\right)  \nonumber \\
&=&D_{osc}^{q/\alpha }(t)C\left( q;\alpha \right) ,  \tag{3.25}
\end{eqnarray}
where $C\left( q;\alpha \right) $is the same as in (3.8).

Numerical simulations of a linear oscillator relaxation include solution of
the Langevin equation (2.19) with an external force $F=-\omega ^2x$ and
calculation of the $q$-th order moments$.$ In Fig. 4 we present the results
obtained for various values of $\omega $ by averaging over 300 realizations,
each of length 4096. The Levy index $\alpha $ is equal 1, and the order of
the moment is 0.25. The values obtained in numerical simulation are depicted
by black points whereas the solid line indicates the values estimated from
Eq. (3.24) by solving FSESE. The vertical mark indicates the value $\tau _x$%
, after which the random process $x(t)$ becomes stationary one. At $t>\tau
_x $ the moment tends to a constant value which is estimated from Eqs.
(3.21) and (3.25). Numerical results demonstrate a good agreement with
theoretical estimates on both non - stationary and stationary stages of
evolution as well.

\section{Solutions to fractional symmetric Fokker - Planck equation}

In this Section, as in previous one, we consider the same examples of
relaxation processes, but governed by FSFPE.

\subsection{Force - free relaxation.}

Setting $F=0$ in Eq. (2.17), we seek for the solution of equation

\begin{equation}
\frac{\partial f}{\partial t}+v\frac{\partial f}{\partial x}=\nu \frac
\partial {\partial v}(vf)+D\frac{\partial ^\alpha }{\partial \left| v\right|
^\alpha }f,  \tag{4.1}
\end{equation}
with an initial condition

\begin{equation}
f\left( x,v,t=0\right) =\delta \left( x-x_{0}\right) \delta (v-v_{0}). 
\tag{4.2}
\end{equation}
For clarity, it is expedient to consider space - homogeneous relaxation at
first.

\subsubsection{Space - homogeneous force - free relaxation.}

We seek for the solution $f(v,t)$ of an equation 
\begin{equation}
\frac{\partial f}{\partial t}=\nu \frac{\partial }{\partial v}(vf)+D\frac{%
\partial ^{\alpha }}{\partial \left| v\right| ^{\alpha }}f  \tag{4.3}
\end{equation}
with an initial condition 
\begin{equation}
f(v,t=0)=\delta (v-v_{0})  \tag{4.4}
\end{equation}
We pass to the characteristic function $\widehat{f}(k,t)$, 
\begin{equation}
f(v,t)=\int_{-\infty }^{\infty }\frac{dk}{2\pi }\widehat{f}%
(k,t)e^{-ikv}\quad ,  \tag{4.5}
\end{equation}
which obeys an equation 
\begin{equation}
\frac{\partial \widehat{f}}{\partial t}+\nu k\frac{\partial \widehat{f}}{%
\partial k}=-D\left| k\right| ^{\alpha }\widehat{f}  \tag{4.6}
\end{equation}
with an initial condition 
\begin{equation}
\widehat{f}\left( k,t=0\right) =e^{ikv_{0}}  \tag{4.7}
\end{equation}
The solution of Eqs. (4.6) and (4.7) is obtained by the method of
characteristics: 
\begin{equation}
\widehat{f}(k,t)=\exp \left\{ ikv_{0}e^{-\nu t}-\left| k\right| ^{\alpha
}D_{ff}^{(v)}(t)\right\}  \tag{4.8}
\end{equation}
where 
\begin{equation}
D_{ff}^{(v)}(t)=\frac{D}{\alpha \nu }(1-e^{-\alpha \nu t})\quad ,  \tag{4.9}
\end{equation}
``ff'' means ``force - free'' case. The space homogeneous relaxation in a
force - free case was first considered in \cite{West2}, where Eq. (4.6) was
obtained and solved with an initial condition (4.7). The process of
relaxation is not a Levy stable processes with independent increments, but,
following terminology used in previous Section, can be named stable - like,
or Levy - like process, since its characteristic function (4.8) contains $%
\left| k\right| ^{\alpha }$ in the exponent, but $D_{ff}^{(v)}(t)$ is not a
linear function of $t$, hence scale parameter $\left( D_{ff}^{(v)}(t)\right)
^{1/\alpha }$ does not scale as $t^{1/\alpha }.$ The stable process with
independent increments arises at small times, 
\begin{equation}
t<<\tau _{v}=1/\alpha \nu \quad ,  \tag{4.10}
\end{equation}
when the exponent in (4.9) can be expanded into power series. With the
accuracy up to linear in $t$ terms inclusively, we get the Levy stable
process. On the other hand, at $t>>\tau _{v}$ the stochastic process $v(t)$
becomes asymptotically stationary process with the stable PDF independent of 
$t$ and with the Levy index $\alpha $ and the scale parameter 
\begin{equation}
\left( D_{ff}^{(v)}(t=\infty )\right) =\left( \frac{D}{\alpha \nu }\right)
^{1/\alpha }\quad .  \tag{4.11}
\end{equation}
In elementary functions the explicit expressions for $f(v,t|v_{0})$ can be
obtained in two cases:

(i) $\alpha =2$ (Brownian space - homogeneous relaxation), for which (e.g., 
\cite{Chandrasekhar}) 
\begin{equation}
f(v,t)=\left[ \frac{2\pi D}{\nu }(1-e^{-2\nu t})\right] ^{-1/2}\exp \left\{ -%
\frac{\nu (v-v_{0}e^{-\nu t})^{2}}{2D(1-e^{-2\nu t})}\right\}  \tag{4.12}
\end{equation}
and (ii) $\alpha =1$ (Cauchy space - homogeneous relaxation), for which 
\begin{equation}
f(v,t)=\frac{D\nu }{\pi }\frac{1-\exp (-\nu t)}{\nu ^{2}(v-v_{0}\exp (-\nu
t))^{2}+D^{2}(1-\exp (-\nu t))^{2}}  \tag{4.13}
\end{equation}

The stationary Maxwell PDF is obtained for $\alpha =2$ only: 
\begin{equation}
f_{st}(v;\alpha =2)=\left( \frac{\nu }{2\pi D}\right) ^{1/2}\exp \left( -%
\frac{\nu v^{2}}{2D}\right) \quad .  \tag{4.14}
\end{equation}

Here it seems expedient to discuss the problems related to stationary
solutions of fractional kinetic equations. In the classical theory of
Brownian motion equilibrium Maxwell PDF over velocity is reached at $%
t\rightarrow \infty .$ It is characterized by the temperature $T$ of
surrounding medium. There exists relation between parameter $D$ of the PDF
of the random source in the Langevin equations, see Eqs. (2.2) - (2.4), and
the friction coefficient $\nu :$%
\begin{equation}
D=\frac{\nu k_BT}m\quad ,  \tag{4.15}
\end{equation}
where $\mathit{m}$ is a mass of Brownian particle, $k_B$ is the Boltzmann
constant. The temperature $T$ is a measure of a mean kinetic energy of the
Brownian particle: 
\begin{equation}
\left\langle E_{kin}\right\rangle =\frac{m\left\langle v^2\right\rangle }2=%
\frac{k_BT}2\quad .  \tag{4.16}
\end{equation}
Equations (4.15) and (4.16) are examples of fluctuation - dissipation
relations. For this case the source $A(t)$ in the Langevin equation is
called a source of internal fluctuations. Relations (4.15) and (4.16) may be
not fulfilled, as it takes place, e.g., in auto-oscillation systems. In such
a case one says that $A(t)$ is a source of external (relatively to the
system considered) fluctuations in Eq. (2.2). However, Maxwell - Boltzmann
exponential form of stationary solutions retains \cite{Klimontovich1}. As to
the Levy motion, the fluctuation - dissipation relations can not be
fulfilled, at least, because of the infinity of square velocity: $%
\left\langle v^2\right\rangle =\infty $ for $0<\alpha <2.$ Therefore, we can
only say about $A(t)$ as about the source of external fluctuations.
Moreover, it stems from the example considered in this subsection, as well
as from the example of linear oscillator considered above, that the
stationary solutions do not possess Maxwell - Boltzmann form but instead
more general form of stable distributions. In present there is no theory of
equilibrium state basing on stable PDFs. Perhaps, the achievement will be
done with the help of Tsallis' statistics and its generalizations, see
recent review \cite{Tsallis} and references therein.

We also write the $q$ - th order fractional moment of the velocity,

\begin{equation}
M_v(t;q,\alpha )=\left\langle \left| v-v_0e^{-\nu t}\right| ^q\right\rangle
=\left( D_{ff}^{(v)}(t)\right) ^{q/\alpha }C(q;\alpha )\quad ,  \tag{4.17}
\end{equation}
where $D_{ff}^{(v)}(t)$ is determined by Eq. (4.9). This formula is compared
with the results of numerical simulations at the end of Subsec.4.

\subsubsection{Space - inhomogeneous force - free relaxation.}

We turn to the force - free relaxation in general case, which is governed by
Eqs. (4.1) and (4.2). We pass to the characteristic function $\widehat{f}%
(\kappa ,k,t)$,

\begin{equation}
f\left( x,v,t\right) =\int_{-\infty }^{\infty }\frac{dx}{2\pi }e^{-i\kappa
x-ikv}\widehat{f}\left( \kappa ,k,t\right) \quad ,  \tag{4.18}
\end{equation}
which obeys an equation

\begin{equation}
\frac{\partial \widehat{f}}{\partial t}+(\nu k-\kappa )\frac{\partial 
\widehat{f}}{\partial k}=-D\left| k\right| ^\alpha \widehat{f},  \tag{4.19}
\end{equation}
with an initial condition

\begin{equation}
\widehat{f}\left( \kappa ,k,t=0\right) =e^{i\kappa x_0+ikv_0}\quad . 
\tag{4.20}
\end{equation}
The solution of Eqs. (4.17), (4.18) is obtained by the method of
characteristics

\begin{eqnarray}
\widehat{f}\left( \kappa ,k,t\right) &=&\exp \{i\kappa (x_0+\frac{v_0}\nu
)+iv_0e^{-\nu t}(k-\frac \kappa \nu )-  \nonumber \\
&&\ \quad \quad -D\int_0^td\tau \left| \frac \kappa \nu +(k-\frac \kappa \nu
)e^{-\nu \tau }\right| ^\alpha \}  \tag{4.21}
\end{eqnarray}
Equation (4.19) admits an analytical inverse Fourier transform in case of
Brownian relaxation,

\begin{equation}
f(x,v,t)=\frac 1{2\pi \Delta ^{1/2}}\exp \left\{ -\frac 1{2\Delta
}(aV^2-2hXV+bX^2)\right\}  \tag{4.22}
\end{equation}
where 
\begin{eqnarray*}
V &=&v-v_0e^{-\nu t}\,, \\
X &=&x-x_0-\frac{v_0}\nu (1-e^{-\nu t})\,, \\
\Delta &=&ab-h^2\,.
\end{eqnarray*}

This result was presented in \cite{Chandrasekhar}.

For an arbitrary $\alpha ,$ $0<\alpha <2,$ the analysis of Eq. (4.19)
becomes rather complicated. However, since we already have an information
about velocity relaxation, we study evolution of a simpler function,

\begin{equation}
f(x,t|x_0,v_0)=\int dvf(x,v,t|x_0,v_0)\quad ,  \tag{4.23}
\end{equation}
whose characteristic function is 
\begin{equation}
\widehat{f}(\kappa ,k=0,t|x_0,v_0)=\exp \left\{ i\kappa x_0+i\kappa \frac{v_0%
}\nu (1-e^{-\nu t})-D_{ff}^{(x)}(t)\left| \kappa \right| ^\alpha \right\}
\quad ,  \tag{4.24}
\end{equation}
where 
\begin{equation}
D_{ff}^{(x)}(t)=\frac D{\nu ^\alpha }\int_0^td\tau (1-e^{-\nu \tau })^\alpha
\quad .  \tag{4.25}
\end{equation}
As well as in previous Sections, we may call the random coordinate in the
process of space - inhomogeneous relaxation as a stable - like process. The
integral in Eq. (4.25) can be evaluated in terms of elementary functions in
two cases,

(i) $\alpha =2$ (Brownian motion), for which 
\[
D_{ff}^{(x)}(t)=\frac D{\nu ^2}\left\{ t-\frac 2\nu (1-e^{-\nu t})+\frac
1{2\nu }(1-e^{-2\nu t})\right\} \quad , 
\]

and (ii) $\alpha =1,$ for which 
\[
D_{ff}^{(x)}(t)=\frac D\nu \left\{ t-\frac{1-e^{-\nu t}}\nu \right\} \quad . 
\]
At large times $t>>\tau _v,$ where $\tau _v$ is defined by Eq. (4.10), we get

\begin{equation}
D_{ff}^{(x)}(t\rightarrow \infty )\rightarrow \frac{Dt}{\nu ^{\alpha }} 
\tag{4.26}
\end{equation}
and the characteristic function (4.24) coincides with the solution of the
Einstein - Smoluchowski equation in a force - free case (with $x_{0}$
substituted by $x_{0}+v_{0}/\nu $), see Eq. (3.5). Thus, in the limit of
large times the random process $x(t)$ becomes $\alpha $ - stable process
with independent increments and with the Levy index $\alpha $ and the scale
parameter $(Dt)^{1/\alpha }/\nu .$ One can see that the space -
inhomogeneous relaxation occurs in two stages, namely, the fast kinetic
stage, at which stationary stable velocity PDF is established after the time
period $\tau _{v}$, and slow diffusion stage, whose characteristic
relaxation time $\tau _{x}$ can be defined, if one introduces an external
scale $L$ of the system considered: 
\begin{equation}
\tau _{x}\approx \frac{(\nu L)^{\alpha }}{D}\text{\quad .}  \tag{4.27}
\end{equation}
For large enough systems $\tau _{x}>>\tau _{v}.$

For the coordinate we also write the $q$ - th order fractional moment which
is estimated in numerical simulations below: 
\begin{equation}
M_{x}(t;q;\alpha )=\left\langle \left| x-x_{0}-\frac{v_{0}}{\nu }(1-e^{-\nu
t})\right| ^{q}\right\rangle =(D_{ff}^{(x)}(t))^{q/\alpha }C(q;\alpha )\quad
.  \tag{4.28}
\end{equation}

Numerical simulation of a force - free relaxation process described by FSFPE
includes the solution of the Langevin equations (2.2) with $F=0$ and
estimation of the moments $M_v(t;q,\alpha )$ and $M_x(t;q,\alpha ).$ The
results are shown in Figs. 5 - 7.\quad

Figures 5 and 6 have an illustrative purpose. In Fig. 5 we show typical
velocity trajectories (at the left) and coordinate trajectories (at the
right) for various values of the Levy index. In the figure $\nu =0.03$ and,
thus the velocity relaxation time $\tau _v$ is equal 20 for $\alpha =1.7,$
26 for $\alpha =1.3$ and 33 for $\alpha =1.0.$ On the main part of the
realizations presented the process $v(t)$ is stationary. In the velocity
realizations large outliers are clearly seen, which appear due to power
asymptotics of the stable PDFs of the velocity. With the Levy index
decreasing ( from top to bottom) the asymptotics become flatter which leads
to the growth of outlier amplitudes. Large outliers of the velocity, in
turn, leads to large jumps on the trajectories $x(t),$ that is, Levy
flights, see illustrations on the right.

In Fig. 6 we depict typical trajectories of the velocity (at the left) and
of the coordinate (at the right) for various values of the friction
coefficient $\nu .$ The Levy index is 1.3. The velocity relaxation time $%
\tau _v$ is equal 8, 26, and 77 for the top, middle and bottom figures,
respectively. For these values the process $v(t)$ is stationary almost on
the whole length of realization. In the figure the difference between
trajectories with different $\tau _v$ are clearly seen: with $\tau _v$
increasing the relaxation of the velocity outliers reduced, whereas
trajectories of $x(t)$ become less cutted up and more smoothed.

In Fig. 7 we depict the velocity moments $M_v$ (at the left) and coordinate
moments $M_x$ (at the right) versus time at different values of friction
coefficient $\nu .$ The moments are obtained by averaging over 50
realizations, each of length 4096. The momentum exponent $q$ is equal 0.25
for the velocity and the coordinate as well, the Levy index is equal 1.3.
The moments estimated by numerical solution of the Langevin equations are
shown by black points, whereas the moments estimated with Eqs. (4.17) and
(4.27) are shown by solid lines. The vertical mark indicates the relaxation
time $\tau _v.$ At the intervals greater than $\tau _v$ the random process $%
v(t)$ becomes stationary and the velocity moment tends to the constant value 
$D/\alpha \nu ,$ see Eq. (4.9). At the same time it follows from right
figures that the process $x(t)$ remains non - stationary one, and the moment
of the coordinate tends to the linear (in a log - log scale) asymptotics,
which has a slope $q/\alpha $ and shown by dotted line in the right figures.
From Fig. 7 we can make a conclusion about agreement between the theoretical
results obtained for a force - free relaxation by solving FSFPE and the
results obtained by numerical solution of the Langevin equations.

\subsection{Relaxation of linear Levy oscillator}

Setting $F=-\omega ^2x$ in Eq. (2.17), we seek for the solution of equation

\begin{equation}
\frac{\partial f}{\partial t}+v\frac{\partial f}{\partial x}-\omega ^{2}x%
\frac{\partial f}{\partial v}=\nu \frac{\partial }{\partial v}(vf)+D\frac{%
\partial ^{\alpha }}{\partial \left| v\right| ^{\alpha }}f\,,  \tag{4.29}
\end{equation}
with an initial condition 
\begin{equation}
f\left( x,v,t=0\right) =\delta \left( x-x_{0}\right) \delta (v-v_{0}). 
\tag{4.30}
\end{equation}
Making Fourier - transform, we get for the characteristic function an
equation

\begin{equation}
\frac{\partial \widehat{f}}{\partial t}+(\nu k-\kappa )\frac{\partial 
\widehat{f}}{\partial k}+\omega ^2k\frac{\partial \widehat{f}}{\partial
\kappa }=-D\left| k\right| ^\alpha \widehat{f}  \tag{4.31}
\end{equation}
with an initial condition 
\begin{equation}
\widehat{f}\left( \kappa ,k,t=0\right) =e^{i\kappa x_0+ikv_0}\,.  \tag{4.32}
\end{equation}
The solution of Eqs. (4.31) and (4.32) is obtained by the method of
characteristics: 
\[
\widehat{f}(\kappa ,k,t)=\exp \{i\omega
^2x_0(-k_1e^{-p_2t}+k_2e^{-p_1t})+iv_0(-p_2k_1e^{-p_2t}+p_1k_2e^{-p_1t})- 
\]
\begin{equation}
-D\int_0^td\tau \left| p_1k_2e^{-p_1\tau }-p_2k_1e^{-p_2\tau }\right|
^\alpha \}  \tag{4.33}
\end{equation}
where 
\[
p_{1,2}=\frac \nu 2\pm \sqrt{\frac{\nu ^2}4-\omega ^2}\,, 
\]
\[
k_{1,2}=\frac 1{p_1-p_2}(k-\frac{p_{1,2}\kappa }{\omega ^2})\,. 
\]
We may introduce 
\[
\nu _1=(\nu ^2-4\omega ^2)^{1/2} 
\]
and rewrite Eq. (4.33) in terms of hyperbolic functions as 
\[
\widehat{f}(\kappa ,k,t)=\exp \{i\omega ^2x_0e^{-\nu t/2}\left[ -\frac{2k}{%
\nu _1}sh\frac{\nu _1t}2+\frac \kappa {\nu _1\omega ^2}\left( \nu sh\frac{%
\nu _1t}2+\nu _1ch\frac{\nu _1t}2\right) \right] + 
\]
\[
+iv_0e^{-\nu t/2}\left[ \frac{2\kappa }{\nu _1}sh\frac{\nu _1t}2-\frac k{\nu
_1}\left( \nu sh\frac{\nu _1t}2-\nu _1ch\frac{\nu _1t}2\right) \right] - 
\]
\begin{equation}
-D\int_0^td\tau e^{-\alpha \nu \tau /2}\left| \frac{2\kappa }{\nu _1}sh\frac{%
\nu _1\tau }2+k\left( ch\frac{\nu _1\tau }2-\frac \nu {\nu _1}sh\frac{\nu
_1\tau }2\right) \right| ^\alpha \}  \tag{4.34}
\end{equation}
This expression is valid for $\nu ^2-4\omega ^2>0\,.$ If $\nu ^2-4\omega
^2<0\,,$then one introduces 
\[
\omega _1=\sqrt{\omega ^2-\frac{\nu ^2}4} 
\]
and makes the following changes in Eq. (4.34): 
\[
ch\frac{\nu _1t}2\rightarrow \cos \omega _1t\,, 
\]
\[
\frac 1{\nu _1}sh\frac{\nu _1t}2\rightarrow \frac 1{2\omega _1}\sin \omega
_1t\,, 
\]
\[
\frac 1{\nu _1}sh\nu _1t\rightarrow \frac 1{2\omega _1}\sin 2\omega _1t\,. 
\]
In aperiodic case, $\nu =4\omega ^2\,,$evidently, 
\[
ch\frac{\nu _1t}2\rightarrow 1\,, 
\]
\[
\frac 1{\nu _1}sh\frac{\nu _1t}2\rightarrow \frac 12t\,, 
\]
\[
\frac 1{\nu _1}sh\nu _1t\rightarrow t\,. 
\]

For the particular case of Brownian oscillator, $\alpha =2,$ one can get the
Gaussian PDF in the phase space by using an inverse Fourier transform of Eq.
(4.34). Further, by using the first and the second derivatives of the
characteristic function at $\kappa =k=0,$ one can get the means and the
variances of the velocity and coordinate as well. For the Brownian
oscillator these formula are presented in \cite{Chandrasekhar}. The
expressions for the means are also valid for all $\alpha $ greater than
unity.

Now we turn to the more complicated fractional moments: 
\[
M_v(q)=\left\langle \left| v\right| ^q\right\rangle
=(D_{osc}^{(v)}(t))^{q/\alpha }C(q;\alpha )\,, 
\]
\begin{equation}
M_x(q)=\left\langle \left| x\right| ^q\right\rangle
=(D_{osc}^{(x)}(t))^{q/\alpha }C(q;\alpha )\,,  \tag{4.35}
\end{equation}
where 
\[
D_{osc}^{(v)}(t)=D\int_0^td\tau e^{-\alpha \nu \tau /2}\left| ch\frac{\nu
_1\tau }2-\frac \nu {\nu _1}sh\frac{\nu _1\tau }2\right| ^\alpha \quad , 
\]
\begin{equation}
D_{osc}^{(x)}(t)=\frac D{\nu ^\alpha }\int_0^td\tau e^{-\alpha \nu \tau
/2}(e^{\nu _1\tau /2}-e^{-\nu _1\tau /2})^\alpha  \tag{4.36}
\end{equation}
and $C(q;\alpha )$ is the same as in previous Sections. Equations (4.35) and
(4.36) are compared with the results of numerical simulation at the end of
Sec. 5.

Equation (4.34), in principle, allows one to study stationary solution,
which is defined by 
\begin{equation}
f_{st}(x,v)=\int_{-\infty }^{\infty }\frac{d\kappa }{2\pi }\int_{-\infty
}^{\infty }\frac{dk}{2\pi }e^{-i\kappa x-ikv}\widehat{f}_{st}(\kappa
,k)\quad ,  \tag{4.37}
\end{equation}
\begin{equation}
\widehat{f}_{st}(\kappa ,k)=\exp \left\{ -D\int_{0}^{\infty }d\tau
e^{-\alpha \nu \tau /2}\left| \frac{2\kappa }{\nu _{1}}sh\frac{\nu _{1}\tau 
}{2}+k\left( ch\frac{\nu _{1}\tau }{2}-\frac{\nu }{\nu _{1}}sh\frac{\nu
_{1}\tau }{2}\right) \right| ^{\alpha }\right\} \quad .  \tag{4.38}
\end{equation}
A simple analytic expression can be obtained for the stationary solution in
the case $\alpha =2$ only: 
\[
\widehat{f}_{st}(\kappa ,k)=\exp \left\{ -\frac{D}{2\nu }\left( \frac{\kappa
^{2}}{\omega ^{2}}+k^{2}\right) \right\} \quad , 
\]
and 
\[
f_{st}(x,v)=\frac{\nu \omega }{2\pi D}\exp \left\{ -\frac{\nu }{2D}%
(v^{2}+\omega ^{2}x^{2})\right\} \quad . 
\]
However, from Eqs. (4.37) and (4.38) one can get some conclusions for
simpler stationary PDFs, namely, 
\[
f_{st}(x)=\int_{-\infty }^{\infty }dvf_{st}(x,v)=\int_{-\infty }^{\infty }%
\frac{d\kappa }{2\pi }e^{-i\kappa x}\widehat{f}_{st}(\kappa ,k=0)\quad , 
\]
and 
\[
f_{st}(v)=\int_{-\infty }^{\infty }dxf_{st}(x,v)=\int_{-\infty }^{\infty }%
\frac{dk}{2\pi }e^{-ikv}\widehat{f}_{st}(\kappa =0,k)\quad . 
\]
Both stationary PDFs are stable ones with the Levy index $\alpha $ and scale
parameters, which are expressed as integrals over $\tau $, see Eq. (4.36).

Though linear oscillator, as we see, admits a strict solution, the general
formulae are not easy to analyze analytically. Therefore, it is instructive
to consider two limit cases, namely, overdamped oscillator and a weakly
damped oscillator. Both cases also are very important in problems related to
a non - linear oscillator.

\section{Limit cases of Levy oscillator}

\subsection{Overdamped Levy oscillator, $\omega /\nu \ll 1.$}

Let us consider the relaxation of the moments in case of an overdamped
oscillator. First we turn to the velocity relaxation. It follows from Eqs.
(4.35) and (4.36), that we can restrict ourselves by zero - order
approximation in $\omega /\nu .$ In this case we get from Eq.(4.36) 
\begin{equation}
D_{osc}^{(v)}(t)\approx \frac D{\alpha \nu }\left( 1-e^{-\alpha \nu \tau
}\right) \quad ,  \tag{5.1}
\end{equation}
which is, of course, coincides with $D_{ff}^{(v)}(t)$ for the force - free
case, and the velocity relaxation time $\tau _v$ for an overdamped
oscillator is given by Eq. (4.10). The conclusion is that the velocity
relaxation for the overdamped oscillator in the main order in small
parameter $\omega /\nu $ is the same as in the force - free case.

Now we consider space relaxation, which differs from the force - free case.
We get from Eq.(4.36) in the first order in $\omega /\nu ,$%
\begin{equation}
D_{osc}^{(x)}(t)\approx \frac{D}{\nu ^{\alpha }}\int_{0}^{t}d\tau \left(
e^{-\omega ^{2}\tau /\nu }-e^{-\nu \tau }\right) ^{\alpha }\quad  \tag{5.2}
\end{equation}
We note that at $\omega =0$ we get Eq. (4.25) describing relaxation in a
force - free case. It follows from Eq. (5.2) that at 
\[
t\gg \tau _{v}=1/\alpha \nu 
\]
the second term in the brackets contributes negligibly, and 
\begin{equation}
D_{osc}^{(x)}(t>>\tau _{v})\approx \frac{D}{\alpha \omega ^{2}\nu ^{\alpha
-1}}\left( 1-e^{-\alpha \omega ^{2}t/\nu }\right) \quad ,  \tag{5.3}
\end{equation}
which coincides with the result obtained in the framework of FSESE, see Eq.
(3.19). Thus we conclude, that at time intervals greater than the velocity
relaxation $\tau _{v},$ the overdamped oscillator can be described with the
help of a more simple kinetic equation, namely, FSESE. For an overdamped
oscillator the relaxation process occurs in two stages: fast velocity
relaxation stage, at which stationary stable velocity PDF is established
during time interval $\tau _{v}$, and slow diffusion stage, at which during
time $\tau _{x\text{ }},$see Eq. (3.20), the stable PDF is established in a
real space.

\subsection{Weakly damped Levy oscillator, $\omega /\nu \gg 1.$}

For this case in the theory of Brownian oscillator there exists the method
of simplifying kinetic description \cite{Klimontovich1}. It is based on the
method of slowly varying amplitudes, or the van-der-Pol method, which is
frequently used, e.g., in radiophysics \cite{Malakhov}, \cite{Akhmanov}. In
this approach the solution of the Langevin equations 
\[
\frac{dx}{dt}=v\,, 
\]
\begin{equation}
\frac{dv}{dt}=-\omega _0^2x-\nu v+A(t)  \tag{5.4}
\end{equation}
is looking in the form 
\[
x=\widetilde{x}\cos \omega _0t+\frac{\widetilde{v}}{\omega _0}\sin \omega
_0t\,, 
\]
\begin{equation}
v=\widetilde{v}\cos \omega _0t-\omega _0\widetilde{x}\sin \omega _0t\quad , 
\tag{5.5}
\end{equation}
where $\widetilde{x}$ and $\widetilde{v}$ are slowly varying (during the
period $2\pi /\omega $ ) amplitudes. The choice of the solution (5.5) is
equivalent to the condition 
\begin{equation}
\cos \omega t\frac{d\widetilde{x}}{dt}+\frac{\sin \omega t}\omega \frac{d%
\widetilde{v}}{dt}=0\text{\quad .}  \tag{5.6}
\end{equation}
We insert (5.5) into (5.4) and, after averaging over the period, get the
Langevin equations for $\widetilde{x}$ and $\widetilde{v}:$%
\[
\frac{d\widetilde{x}}{dt}+\frac \nu 2\widetilde{x}=A_{\widetilde{x}}(t)\,, 
\]
\begin{equation}
\frac{d\widetilde{v}}{dt}+\frac \nu 2\widetilde{v}=A_{\widetilde{v}}(t)\,, 
\tag{5.7}
\end{equation}
where 
\[
A_{\widetilde{x}}(t)=-\frac 1{2\pi }\int_{t-2\pi /\omega }^tdt^{\prime
}A(t^{\prime })\sin \omega t^{\prime }\quad , 
\]
\begin{equation}
A_{\widetilde{v}}(t)=\frac \omega {2\pi }\int_{t-2\pi /\omega }^tdt^{\prime
}A(t^{\prime })\cos \omega t^{\prime }\quad .  \tag{5.8}
\end{equation}
It follows from (5.7) that the Langevin sources $A_{\widetilde{x}}(t)$ and $%
A_{\widetilde{v}}(t)$ does not contain ``fast'' time $2\pi /\omega .$
Therefore it stems from Eq. (5.8) that $A(t)$ can be represented as 
\begin{equation}
A(t)\approx a(t)\cos \omega t-b(t)\sin \omega t\quad ,  \tag{5.9}
\end{equation}
where $a(t)$ and $b(t)$ are random stationary functions which are related to 
$A_{\widetilde{x}}(t)$ and $A_{\widetilde{v}}(t)$ as 
\[
a(t)=2A_{\widetilde{v}}(t)\text{\quad ,} 
\]
\begin{equation}
b(t)=-2\omega A_{\widetilde{x}}(t)\quad .  \tag{5.10}
\end{equation}
Equations (5.9) and (5.10) have the following meaning \cite{Malakhov}.
According to Eqs. (5.7) and (5.8) the random force influences oscillator by
means of slowly varying components $A_{\widetilde{x}}(t)$ and $A_{\widetilde{%
v}}(t)$ (or $a(t)$ and $b(t)$) only. Therefore, if one considers the random
influence on the weakly damped oscillator, the main components of the random
force are singled out by Eq. (5.9). Further, if $A(t)$ is a stationary
Gaussian process, then after constructing expression for the correlation
function of this process, one can convince himself that one - point PDFs of $%
A(t)$, $a(t)$ and $b(t)$ coincide \cite{Akhmanov}. We assume that the
conclusion about identical PDFs of $A(t)$, $a(t)$ and $b(t)$ is also valid
for (symmetrical) stable PDFs notwithstanding the fact that the proof of
this statement is not so trivial as in the Gaussian case. It follows from
this coincidence that the PDFs of the processes 
\[
L(\Delta t)=\int_t^{t+\Delta t}dt^{\prime }A(t^{\prime })\text{\quad ,} 
\]
\[
L_a(\Delta t)=\int_t^{t+\Delta t}dt^{\prime }a(t^{\prime })\quad , 
\]
\begin{equation}
L_b(\Delta t)=\int_t^{t+\Delta t}dt^{\prime }b(t^{\prime })  \tag{5.11}
\end{equation}
also coincide, and thus, with taking Eq.(2.5), the PDFs of $L_a$ , $L_b$ are 
\begin{equation}
w(L_{a,b})=\int_{-\infty }^\infty \frac{dk}{2\pi }\exp \left(
-ikL_{a,b}-D\left| k\right| ^\alpha \Delta t\right) \quad .  \tag{5.12}
\end{equation}
We also define the processes 
\[
L_{\widetilde{v}}(\Delta t)=\int_t^{t+\Delta t}dt^{\prime }A_{\widetilde{v}%
}(t^{\prime })\quad , 
\]
\begin{equation}
L_{\widetilde{x}}(\Delta t)=\int_t^{t+\Delta t}dt^{\prime }A_{\widetilde{x}%
}(t^{\prime })\quad ,  \tag{5.13}
\end{equation}
which, according to (5.10) are related to $L_a,$ $L_b$ as 
\[
L_a(\Delta t)=2L_{\widetilde{v}}(\Delta t)\quad , 
\]
\begin{equation}
L_b(\Delta t)=-2\omega L_{\widetilde{x}}(\Delta t)\quad .  \tag{5.14}
\end{equation}
Now, with the help of Eqs. (5.12) and (5.14) we are able to get the
characteristic functions $\widehat{w}(L_{\widetilde{x}})$ and $\widehat{w}%
(L_{\widetilde{v}})$ as well as their PDFs $w(L_{\widetilde{x}})$ and $w(L_{%
\widetilde{v}})$ $:$%
\[
\widehat{w}(L_{\widetilde{x}})=\exp \left( -D_{\widetilde{x}}\left| k\right|
^\alpha \Delta t\right) \quad , 
\]
\begin{equation}
\widehat{w}(L_{\widetilde{v}})=\exp \left( -D_{\widetilde{v}}\left| k\right|
^\alpha \Delta t\right) \quad ,  \tag{5.15}
\end{equation}
where 
\begin{equation}
D_{\widetilde{x}}=\frac D{(2\omega )^\alpha }\quad ,D_{\widetilde{v}}=\frac
D{2^\alpha }\quad .  \tag{5.16}
\end{equation}

Equation for $f(\widetilde{x},\widetilde{v},t)$ is derived in the manner
analogous to FSESE. An initial equation is 
\begin{eqnarray}
f(\widetilde{x},\widetilde{v},t+\Delta t) &=&\int \int d(\Delta \widetilde{x}%
)d(\Delta \widetilde{v})f(\widetilde{x}-\Delta \widetilde{x},v-\Delta 
\widetilde{v},t)\times  \nonumber \\
&&\quad \quad \quad \times \Psi (\widetilde{x}-\Delta \widetilde{x},v-\Delta 
\widetilde{v};\Delta \widetilde{x},\Delta \widetilde{v},\Delta t)\quad , 
\tag{5.17}
\end{eqnarray}
where $\Psi $ is the transition probability. For the increments $\Delta 
\widetilde{x},$ $\Delta \widetilde{v}$ we get from Eqs. (5.7) 
\[
\Delta \widetilde{x}+\frac{\nu }{2}\widetilde{x}\Delta t=L_{\widetilde{x}%
}(\Delta t)\quad , 
\]
\begin{equation}
\Delta \widetilde{v}+\frac{\nu }{2}\widetilde{v}\Delta t=L_{\widetilde{v}%
}(\Delta t)\quad ,  \tag{5.18}
\end{equation}
where the PDFs for $L_{\widetilde{x}}$ and $L_{\widetilde{v}}$ are given by
Eqs. (5.15) and (5.16). Now we construct $\Psi .$ From the structure of Eqs.
(5.18) it follows that 
\begin{equation}
\Psi (\widetilde{x}-\Delta \widetilde{x},v-\Delta \widetilde{v};\Delta 
\widetilde{x},\Delta \widetilde{v},\Delta t)=\Psi _{\widetilde{x}}(%
\widetilde{x}-\Delta \widetilde{x};\Delta \widetilde{x},\Delta t)\Psi _{%
\widetilde{v}}(\widetilde{v}-\Delta \widetilde{v};\Delta \widetilde{v}%
,\Delta t)\quad ,  \tag{5.19}
\end{equation}
where 
\[
\Psi _{\widetilde{x}}(\widetilde{x};\Delta \widetilde{x},\Delta
t)=\int_{-\infty }^{\infty }\frac{d\kappa }{2\pi }\exp \left( -i\kappa
\left( \Delta \widetilde{x}+\frac{\nu }{2}\widetilde{x}\Delta t\right) -D_{%
\widetilde{x}}\left| \kappa \right| ^{\alpha }\Delta t\right) \quad , 
\]
\begin{equation}
\Psi _{\widetilde{v}}(\widetilde{v};\Delta \widetilde{v},\Delta
t)=\int_{-\infty }^{\infty }\frac{dk}{2\pi }\exp \left( -i\kappa \left(
\Delta \widetilde{v}+\frac{\nu }{2}\widetilde{v}\Delta t\right) -D_{%
\widetilde{v}}\left| k\right| ^{\alpha }\Delta t\right) \quad .  \tag{5.20}
\end{equation}
Insert Eqs.(5.19) and (5.20) into Eq. (5.17), expand into power series in $%
\Delta t$ and tend $\Delta t$ to zero. As the result we get 
\[
\frac{\partial f}{\partial t}=-\int \int d(\Delta \widetilde{x})d(\Delta 
\widetilde{v})f(\widetilde{x}-\Delta \widetilde{x},v-\Delta \widetilde{v}%
,t)\int_{-\infty }^{\infty }\frac{d\kappa }{2\pi }\int_{-\infty }^{\infty }%
\frac{dk}{2\pi }\exp \left( -i\kappa \Delta \widetilde{x}-ik\Delta 
\widetilde{v}\right) \times 
\]
\begin{equation}
\left[ i\kappa \frac{\nu }{2}(\widetilde{x}-\Delta \widetilde{x})+ik\frac{%
\nu }{2}(v-\Delta \widetilde{v})+D_{\widetilde{x}}\left| \kappa \right|
^{\alpha }+D_{\widetilde{v}}\left| k\right| ^{\alpha }\right] \quad . 
\tag{5.21}
\end{equation}
Transforming the terms in the right hand side as it is described in Sec. 2,
we arrive at the differential equation for $f(\widetilde{x},\widetilde{v},t)$
:

\begin{equation}
\frac{\partial f}{\partial t}=\frac{\nu }{2}\frac{\partial }{\partial 
\widetilde{x}}(\widetilde{x}f)+\frac{\nu }{2}\frac{\partial }{\partial 
\widetilde{v}}(\widetilde{v}f)+D_{\widetilde{x}}\frac{\partial ^{\alpha }f}{%
\partial \left| \widetilde{x}\right| ^{\alpha }}+D_{\widetilde{v}}\frac{%
\partial ^{\alpha }f}{\partial \left| \widetilde{v}\right| ^{\alpha }}\quad .
\tag{5.22}
\end{equation}
We find the solution of Eq. (5.22) with an initial condition 
\begin{equation}
f(\widetilde{x},\widetilde{v},t=0)=\delta (\widetilde{x}-x_{0})\delta (%
\widetilde{v}-v_{0})\quad .  \tag{5.23}
\end{equation}
We pass to the characteristic function $\widehat{f}(\kappa ,k,t),$%
\begin{equation}
f(\widetilde{x},\widetilde{v},t)=\int_{-\infty }^{\infty }\frac{d\kappa }{%
2\pi }\int_{-\infty }^{\infty }\frac{dk}{2\pi }\exp (-i\kappa \widetilde{x}%
-ik\widetilde{v})\widehat{f}(\kappa ,k,t)\quad  \tag{5.24}
\end{equation}
which obeys an equation 
\begin{equation}
\frac{\partial \widehat{f}}{\partial t}=-\frac{\nu }{2}\kappa \frac{\partial 
\widehat{f}}{\partial \kappa }-\frac{\nu }{2}k\frac{\partial \widehat{f}}{%
\partial k}-D_{\widetilde{x}}\left| \kappa \right| ^{\alpha }\widehat{f}-D_{%
\widetilde{v}}\left| k\right| ^{\alpha }\widehat{f}\quad ,  \tag{5.25}
\end{equation}
with an initial condition 
\begin{equation}
\widehat{f}(\kappa ,k,t=0)=e^{i\kappa x_{0}+ikv_{0}}\quad .  \tag{5.26}
\end{equation}
The solution of Eqs. (5.25) and (5.26) is 
\begin{eqnarray}
\widehat{f}(\kappa ,k,t) &=&\exp \{i\kappa x_{0}e^{-\nu t/2}+ikv_{0}e^{-\nu
t/2}-\frac{2D_{\widetilde{x}}}{\alpha \nu }\left| \kappa \right| ^{\alpha
}(1-e^{-\alpha \nu t/2})  \nonumber \\
&&\ \quad \quad -\frac{2D_{\widetilde{v}}}{\alpha \nu }\left| k\right|
^{\alpha }(1-e^{-\alpha \nu t/2}).\text{ }  \tag{5.27}
\end{eqnarray}
We also get the fractional moments: 
\[
M_{\widetilde{v}}(t;q,\alpha )=\left\langle \left| \widetilde{v}%
-v_{0}e^{-\nu t/2}\right| ^{q}\right\rangle =(D_{osc}^{(\widetilde{v}%
)}(t))^{q/\alpha }C(q;\alpha )\,, 
\]
\begin{equation}
M_{\widetilde{x}}(t;q,\alpha )=\left\langle \left| \widetilde{x}%
-x_{0}e^{-\nu t/2}\right| ^{q}\right\rangle =(D_{osc}^{(\widetilde{x}%
)}(t))^{q/\alpha }C(q;\alpha )\,,  \tag{5.28}
\end{equation}
where 
\[
D_{osc}^{(\widetilde{v})}=\frac{2D_{\widetilde{v}}}{\alpha \nu }\left(
1-e^{-\alpha \nu t/2}\right) \,, 
\]
\begin{equation}
D_{osc}^{(\widetilde{x})}=\frac{2D_{\widetilde{x}}}{\alpha \nu }\left(
1-e^{-\alpha \nu t/2}\right)  \tag{5.29}
\end{equation}
It follows from the last equations that, opposite to the case of overdamped
Levy oscillator, the relaxation times for the weakly damped oscillator are
equal, 
\begin{equation}
\tau _{v}=\tau _{x}=\tau =\frac{2}{\alpha \nu }\,\quad ,  \tag{5.30}
\end{equation}
and, thus for a weakly damped oscillator it is impossible to distinguish
kinetic and diffusion stages of relaxation. At time intervals greater than $%
\tau $ the random processes $\widetilde{x}(t)$ and $\widetilde{v}(t)$
becomes stationary ones with stable PDFs. The characteristic function of the
stationary state is defined from Eq. (5.27): 
\[
\widehat{f}_{st}(\kappa ,k)=\exp \left( -\frac{2D_{\widetilde{x}}}{\alpha
\nu }\left| \kappa \right| ^{\alpha }-\frac{2D_{\widetilde{v}}}{\alpha \nu }%
\left| k\right| ^{\alpha }\right) \quad , 
\]
and the PDF retains Maxwell - Boltzmann form for $\alpha =2$ only.

Numerical simulations of a linear oscillator relaxation include solution the
Langevin equations (2.2) with an external force $F=-\omega ^2x$ and
subsequent calculation of the velocity and coordinate moments. The results
are shown in Figs. 8 - 13.

Figures 8 - 10 has an illustrative character. In Fig. 8 the typical
trajectories are depicted of the velocity (at the top) and of the coordinate
(at the bottom) for the overdamped oscillator (at the left) and for the
weakly damped oscillator (at the right), respectively. The frequency value
is equal 0.003 and 0.3 for the overdamped and weakly damped oscillators,
respectively. The friction coefficient is equal 0.03 and the Levy index is
equal 1.3. In the figures the trajectories are shown which have a single
large outlier. It allows us to demonstrate visually the difference in
behaviors of two kinds of oscillator: the relaxation process for an
overdamped oscillator (at the left) resembles relaxation in a force - free
case and is radically different from rapidly oscillating behaviour of the
velocity and coordinate of a weakly damped oscillator (at the right).

In Fig. 9 trajectories on the phase plane $(x,v)$ are depicted for (a)
weakly damped oscillator, $\omega =0.06$, and (b) overdamped oscillator, $%
\omega =0.009,$ respectively. The friction coefficient $\nu $ is 0.02, and
the Levy index $\alpha $ is 1.2. In simulations a single trajectory having
the length 12.000 is used. In both cases the trajectory starts from the
point (0, 0), and, then is ``thrown away'' from this point by Levy
``pushes'' produced by external source in the Langevin equation (2.2). For
the weakly damped oscillator the whole picture is a set of spirals gathering
in a focus at (0, 0). For the overdamped oscillator the phase trajectories
are a set of curves gathering in a node at (0, 0) without circumvolution
around it. In Fig. 10 the trajectories are depicted on the plane $(x,E)$
where $E$ is the energy of oscillator, $E=v^2/2+(\omega x)^2/2,$ for the two
Levy indexes $\alpha =1.9$ (above) and $\alpha =1.6$ (below). In simulations
we use the following parameters: $\omega =0.003,\nu =0.002,$ the length of
trajectory is 4096. In the top figure the large jumps are almost absent, and
the trajectory resembles that of the Brownian oscillator. In the bottom
figure a single large jump, that is ``Levy flight'' occurs, which is due to
slowly decreasing power law asymptotics of stable distribution.

In Fig. 11 we show the velocity moments $M_v$ (in the top) and coordinate
moments $M_x$ (in the bottom) versus $t$ for the overdamped oscillator (at
the left) and for the weakly damped oscillator (at the right),
correspondingly. The oscillation frequencies $\omega $ are equal 0.01 and
0.1 for the overdamped and the weakly damped oscillators, correspondingly,
the friction coefficient is equal 0.03. The order of the moment is 0.5, and
the Levy index is 1.3. The moments obtained by numerical simulation are
shown by black points, whereas the theoretical values, see Eqs. (4.35) and
(4.36), are shown by solid line. The numerical values are obtained by
averaging over 200 realizations, each of length 1024. The vertical marks
indicate the velocity relaxation time $\tau _v=1/\alpha \nu $ and coordinate
relaxation time $\tau _x=\nu /\alpha \omega ^2$ for the overdamped
oscillator, correspondingly, as well as relaxation time $\tau _v=\tau
_x=2/\alpha \nu $ for the weakly damped oscillator. Both theoretical and
numerical curves reach ``plateau'' at the intervals greater than the
relaxation times. It implies that the processes $v(t)$ and $x(t)$ become
stationary ones. The figures demonstrate an important difference between
overdamped and weakly damped oscillators: for the overdamped oscillator
coordinate relaxation time is much greater than the velocity relaxation
time, whereas for the weakly damped oscillator both times are the same. From
Fig. 11 we also make a conclusion about good quantitative agreement between
theoretical results obtained by solving FSFPE and numerical solution of the
corresponding Langevin equations. In this respect it is worthwhile to point
ones attention to coincidence of theoretical curve and numerical dependence
at nonstationary parts.

In order to estimate the influence of power asymptotics on the evolution of
the moments, we replace stable PDFs in the Langevin equation by
``truncated'' ones, that is those, in which the large values of random
quantities are cutted off. In Fig. 12 the velocity moments versus time are
depicted for the weakly damped oscillator in a log - log scale. The
oscillator frequency is 0.05, the friction coefficient is 0.01, the order of
the moment is 0.25, and the Levy index is 1.3. The moments are obtained by
averaging over 1500 realizations, each of length 2048, thus the total number
of points is $3\cdot 10^{6}.$ The mode of maximum value (that is, the most
probable value) is of the order $N^{1/\alpha }.$ In the figure the moments
obtained numerically are shown by black points. The Langevin source $A(t)$
is modelled as a consequence of independent random variables possessing
truncated stable PDF, that is $\left| A(t)\right| <A_{\max }=1600;300;50$
for the cases a; b; c, respectively. The solid line indicates the moments
estimated analytically from the FSFPE. It is seen, that the role of large
outliers increases with time increasing, since large values become more and
more probable. Therefore, the discrepancy between theoretical results and
numerical simulations using truncated PDF grows with time growing. With the
truncation parameter decreasing the discrepancy increases. Thus, as it is
clearly seen, the discrepancies are most essential at the stationary stage
of evolution.

We have already mentioned that our studies demonstrate a good quantitative
agreement between theory based on FSFPE and numerical simulations based on
the Langevin equations. In order to show this fact more precisely, we
estimated velocity and coordinate moments by averaging over 50$\cdot 10^3$
realizations with the total number of points $10^8,$ which is much larger
than in simulations presented in Fig.11. Here we use the following
parameters: $\omega =0.05,\nu =0.01,q=0.25,\alpha =1.3$. By black points the
results of simulation are shown, whereas the solid lines indicate
theoretical values estimated with using Eqs. (4.35) and (4.36). The
theoretical values of $\tau _v$ and $5\tau _v$ are indicated by arrows. It
is seen that numerical results strictly repeat all bends of theoretical
curves at the non -stationary stage of evolution.

\section{Results}

A large section of statistical physics deals with evolution of the systems
influenced by random Gaussian forces. In this paper we study linear
relaxation of the systems influenced by random forces which are distributed
with symmetric stable probability laws. Stable laws (as the Gaussian one)
are the limit ones for probability laws for sums of independent identically
distributed random variables. Therefore, they appear in problems, whose
result is determined by the sum of a great number of independent identical
factors.

The main results are as follows.

1. We get fractional symmetric Fokker - Planck equation (FSFPE) and
fractional symmetric Einstein - Smoluchowski equation (FSESE). They
generalize the Fokker - Planck and Einstein - Smoluchowski equations for the
Brownian motion. The FSFPE describes a linear relaxation in the phase space
of the systems influenced by stochastic forces distributed with symmetric
stable law. The FSFPE contains fractional velocity derivative instead of the
second one. The FSESE describes relaxation in a real space. It contains
fractional space derivative. To get these equations we use an approach
analogous to that used by Chandrasekhar in order to derive the kinetic
equations for the Brownian motion.

We restrict ourselves by the case of one - dimensional symmetric stable
PDFs. The approach can be generalized in order to get the kinetic equations
for two - and three - dimensional real space as well as for the case of
asymmetric stable PDFs of stochastic external forces. At these ways the
problems appear, which are connected with, at first, many - dimensional
generalization of stable probability laws and, at the second, with many -
dimensional generalization of fractional derivatives. On the other hand, the
case of asymmetric stable PDFs may appear to be important for the
description of asymmetric diffusion. These problems in relation with the
kinetic equations are discussed in some recent papers \cite{Chaives}, \cite
{Zolotarev}.

2. With the help of kinetic equations obtained we consider the processes of
a linear relaxation for two problems: force - free relaxation and relaxation
of a linear oscillator. We get general analytic solutions of FSFPE and
FSESE, and expressions for fractional velocity and coordinate moments as
well.

3. For the both problems we solve numerically Langevin equations with the
random source, which is a discrete approximation to a white Levy noise.
After averaging over many realizations we estimate fractional moments and
compare numerical results with the results of analytical solutions to the
kinetic equations. Analytical and numerical results appear to be in a
quantitative agreement.

4. The process of a linear relaxation in a force - free case can be divided
into two stages which are described in the framework of FSFPE:\ ``fast''
stage, at which stationary stable PDF over velocity is established, and
``slow'' diffusion stage, at which relaxation in a real space occurs. The
latter process can be described asymptotically as the Levy stable process
with independent increments. The characteristic time of velocity relaxation
is $\tau _{v}=1/\alpha \nu ,$ where $\nu $ is the friction coefficient in
the Langevin equation, and $\alpha $ is the Levy index of stable PDF of the
random force in the Langevin equation. The characteristic time of relaxation
in a real space is $\tau _{x}=(\nu L)^{\alpha }/D,$ where $L$ is an external
size of the system considered, and $D^{1/\alpha }$ is the scale parameter of
the PDF of the stochastic force. For the large enough systems $\tau
_{x}>>\tau _{v},$ and division of relaxation process into two stages is
valid. For this case at the diffusion stage relaxation can be described in
the framework of FSESE, which describes anomalous superdiffusion, that is,
the diffusion process when the particle displacement grows as $t^{1/\alpha
}, $ $0<\alpha <2.$

5. When studying relaxation of a linear oscillator, it is expedient to
distinguish between two cases:\ overdamped oscillator, $\omega /\nu <<1,$
and weakly damped oscillator, $\omega /\nu >>1.$ Both cases are of special
importance in nonlinear generalizations of the theory presented. We study in
detail, both analytically and numerically, both limit cases, and point
attention to substantially different properties of relaxation processes in
two cases.

6. The relaxation of an overdamped oscillator occurs in two stages, which
are described in the framework of FSFPE: ``fast'' stage, at which during
time interval $\tau _{v}=1/\alpha \nu $ stationary stable PDF over velocity
is established, and ``slow'' diffusion stage, at which during the time
interval $\tau _{x}=\nu /\alpha \omega ^{2}$ stationary stable PDF in a real
space is established. At the diffusion stage the relaxation of an overdamped
oscillator can be described in the framework of FSESE.

7. It is known that for a weakly damped Brownian oscillator a method occurs,
which allows one to simplify kinetic description by passing to slowly
varying (at the period of oscillations) random variables. We generalize this
approach to the case of a weakly damped Levy oscillator and get kinetic
equation for the PDF depending on slow variables. This equation generalizes
the kinetic equation for a weakly damped Brownian oscillator and contains
fractional derivatives over velocity and coordinate. Its structure is more
simple than that of FSFPE, therefore, it can be applied for non - linear
problems. We find stationary solution of the obtained equation and show that
relaxation process can not be divided into two steps. Both velocity and
coordinate relax during the same time interval $\tau =2/\alpha \nu .$

\section{Acknowledgements}

The authors thank F. Mainardi (Bologna University) for valuable comments on
problems related to fractional calculus and diffusion - like equations. This
paper was supported by the Project ``Chaos - 2'' of National Academy of
Science of Ukraine and by the INTAS Project 98 - 01.

\newpage FIGURE CAPTIONS.

Fig.1. $C^{1/q}(q;\alpha )$ versus $q$ for different $\alpha
=1.0;1.3;1.6;1.9.$

Fig.2. Force - free relaxation in the framework of FSESE. The moment $M_x$%
versus $t$ at different values of moment exponent $q$. The Levy index $%
\alpha $ is 1.

Fig.3. Force - free relaxation in the framework of FSESE. The exponent $\mu $
of the time dependence of the moment $M_x,$ see Eq. (3.13), versus the Levy
index $\alpha $. The moment order is $q=0.25.$ Theoretical dependence 0.25/$%
\alpha $ is depicted by dotted line. Numerical results are depicted by black
points.

Fig.4. Relaxation of a linear oscillator in the framework of FSESE. The $q$
- th order coordinate moment versus time in a log - log scale. The results
of numerical simulations are depicted by black points, the moment obtained
from FSESE is shown by solid line. Vertical marks indicate the time of
coordinate relaxation.

Fig.5. Force - free relaxation in the framework of FSFPE. The typical
trajectories of the velocity (at the right) and of the coordinate (at the
left) for various values of the Levy index $\alpha .$ The friction
coefficient $\nu $ is 0.03.

Fig.6. Force - free relaxation in the framework of FSFPE. The typical
trajectories of the velocity (at the left) and of the coordinate (at the
right) for various values of the friction coefficient $\nu .$ The Levy index
is 1.3.

Fig.7. Force - free relaxation in the framework of FSFPE. The velocity
moments (at the left) and the coordinate moments (at the right) versus time
in a log - log scale. Black points indicate results of numerical simulation
of the Langevin equations, the solid lines indicates the moments obtained
from FSFPE. The moment order $q$ is 0.25, The Levy index $\alpha $ is 1.3.
Horizontal dots show the time $\tau _v$ of the velocity relaxation and the
time $5\tau _v.$ The dotted lines at the right figures indicate theoretical
values $q/\alpha $ of straight - line asymptotics (in a log - log scale) for
the moment of the coordinate.

Fig.8. Relaxation of linear oscillator in the framework of FSFPE. Typical
trajectories of the velocity (top) and the coordinate (bottom) for
overdamped (left) and weakly damped (right) oscillators. The frequencies are
0.003 (overdamped) and 0.3 (weakly damped) correspondingly, the friction
coefficient is 0.03, the Levy index is 1.3.

Fig.9. Trajectories on the phase plane for (a) weakly damped, and (b)
overdamped oscillators, respectively. The parameters are $\omega =0.06$ and $%
\omega =0.009$ for (a) and (b), respectively, $\nu =0.02$, and $\alpha =1.2.$

Fig.10. Trajectories on the plane ($x$, $E$), $E$ is the oscillator energy,
for two Levy indexes $\alpha =1.9$ (above) and $\alpha =1.6$ (below),
respectively. The parameters are $\omega =0.003,$ $\nu =0.002$.

Fig.11. Relaxation of linear oscillator in the framework of FSFPE. The
velocity moments (at the top) and the coordinate moments (at the bottom)
versus time in a log - log scale. At the left the results for the overdamped
oscillator are presented, $\omega =0.01,$ whereas at the right the results
for the weakly damped oscillator are presented, $\omega =0.1.$ The frequency
coefficient $\nu $ is 0.03, the order $q$ of the moments is 0.25, the Levy
index $\alpha $ is 1.3. Black points indicate results of numerical
simulation of the Langevin equations, the solid lines indicates the moments
obtained from FSFPE, see Eqs. (4.35) and (4.36). The arrows show velocity
and coordinate relaxation times.

Fig.12. Relaxation of linear oscillator in the framework of FSFPE. The
velocity moments versus time in a log - log scale for different truncation
parameters $A_{\max }$ for the stable PDFs of the noise in the Langevin
equations. The parameters used in simulations are as follows: $\omega
=0.05,\nu =0.01,q=0.25,\alpha =1.3.$ $A_{\max }$ is 1.500, 300 and 50 for
cases $a$,$b$,$c$, respectively.

Fig.13. Relaxation of linear oscillator in the framework of FSFPE. The
velocity moments (top) and coordinate moments (bottom) versus $t$ in a log -
log scale. The parameters of the simulations are as follows: $\omega
=0.05,\nu =0.01,q=0.25,\alpha =1.3.$ The moments are estimated by averaging
over 50$\cdot 10^3$ realizations each of the length 2048. The vertical marks
indicate relaxation times $\tau _v,\tau _x$ and 5$\tau _v,5\tau _x,$ as well.

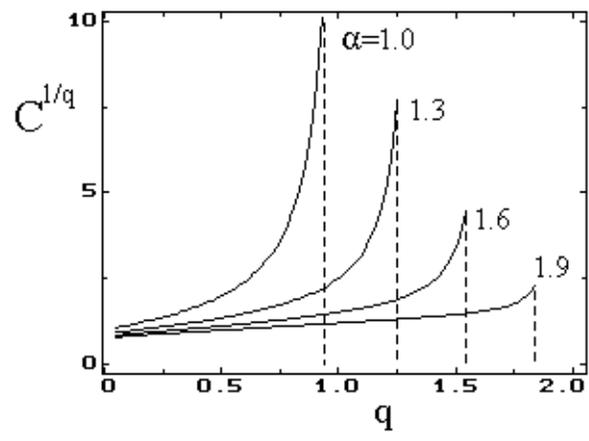
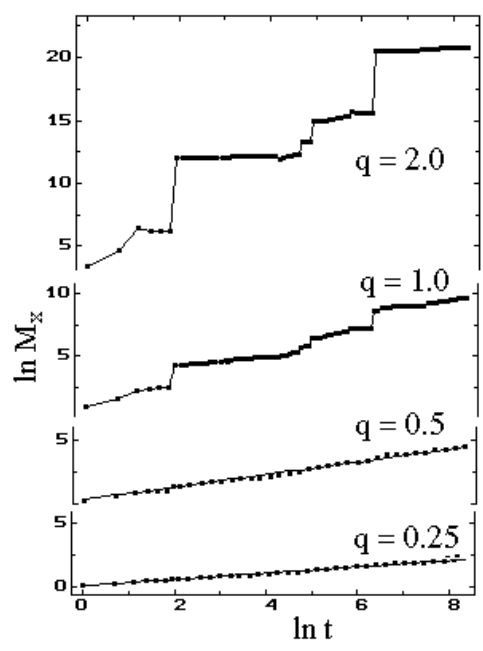
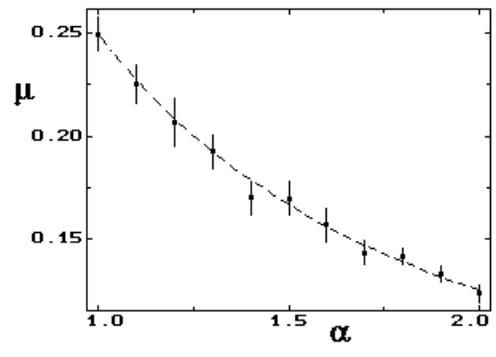
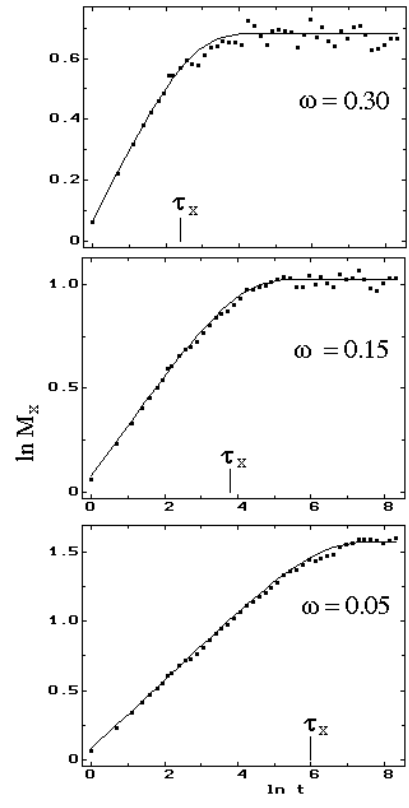
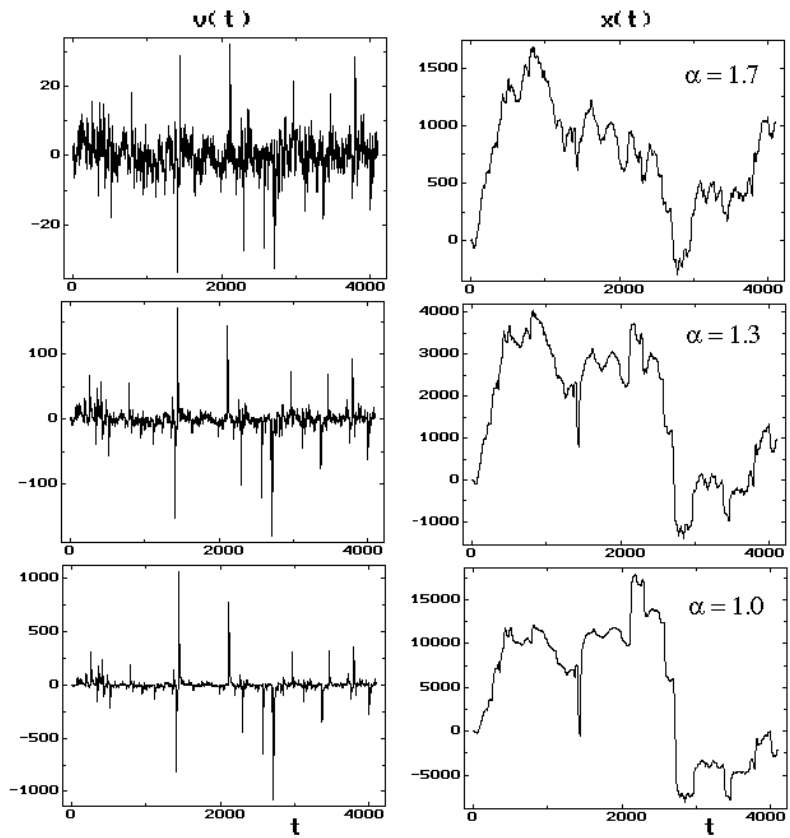
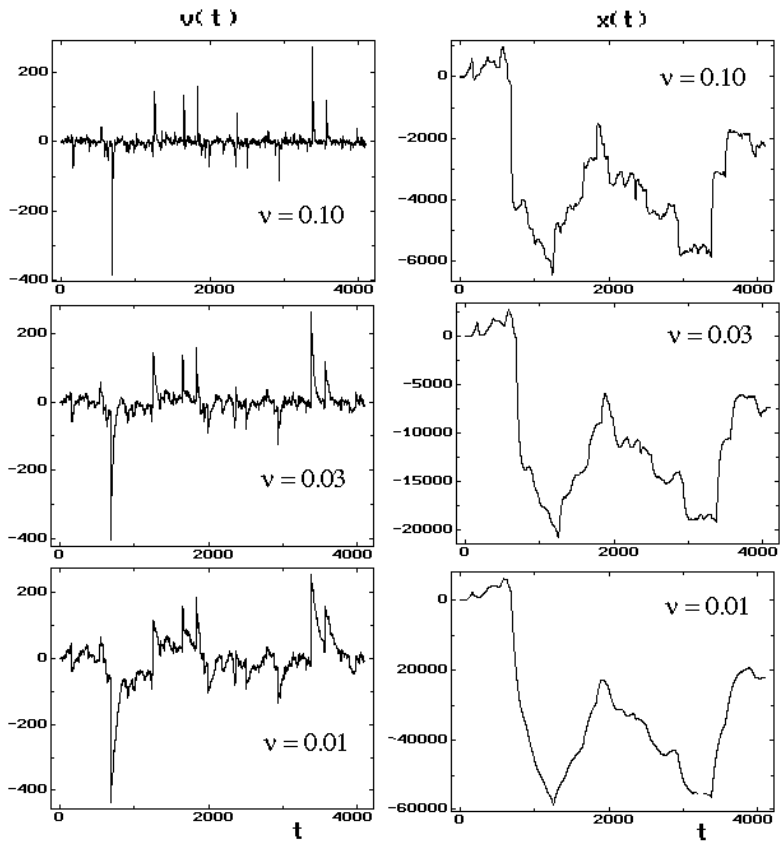
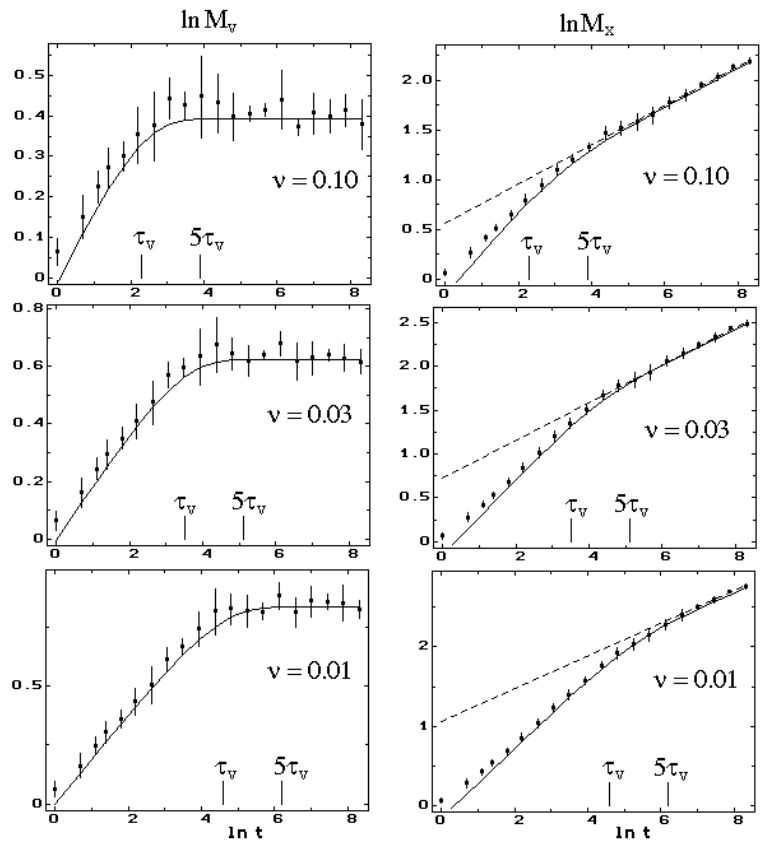
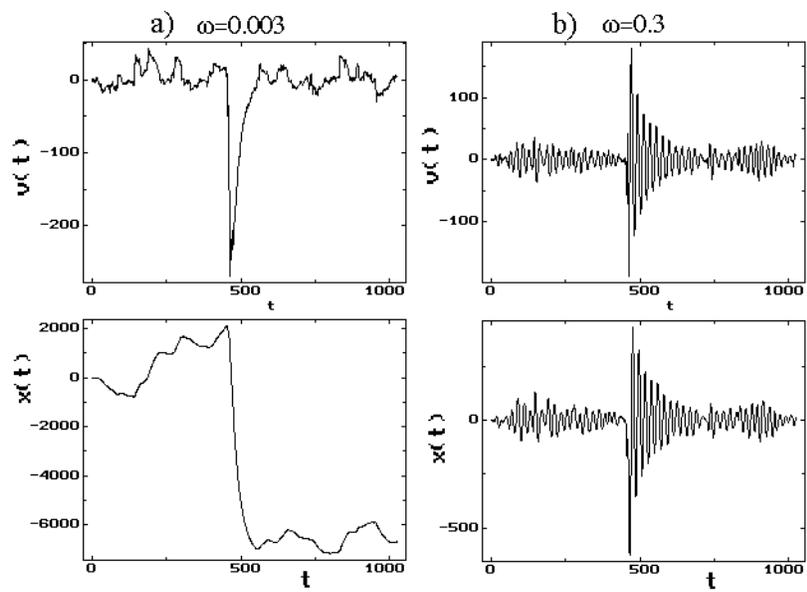
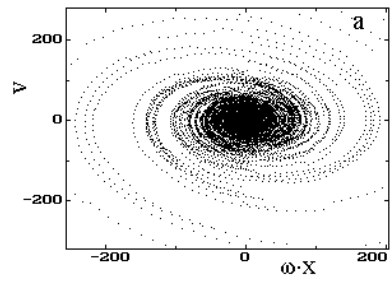
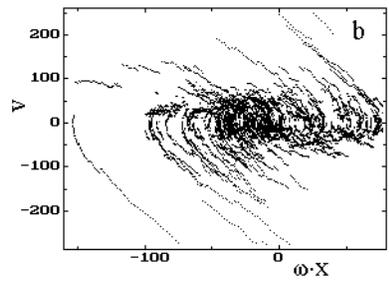
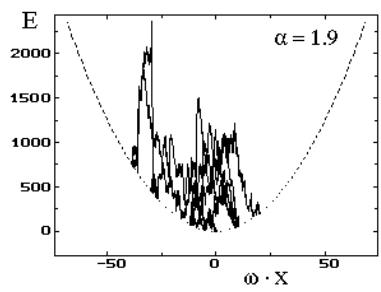
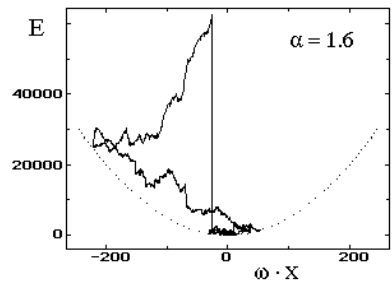
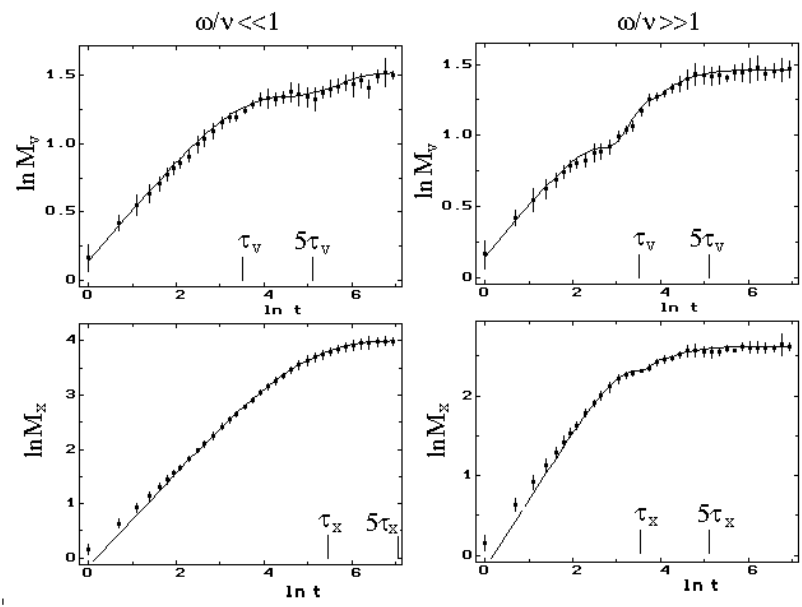
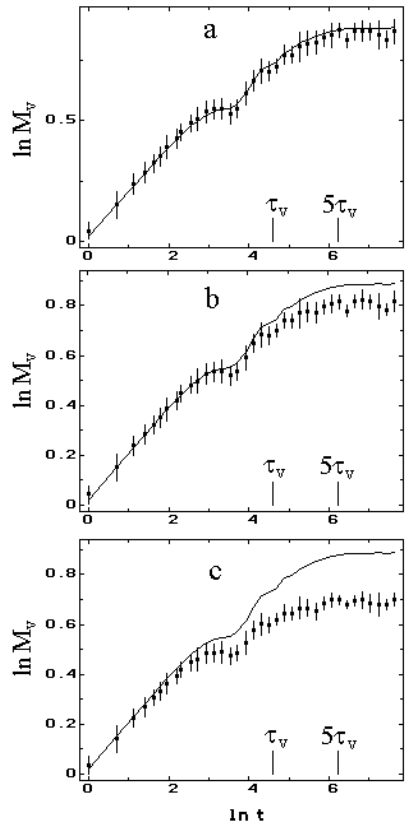
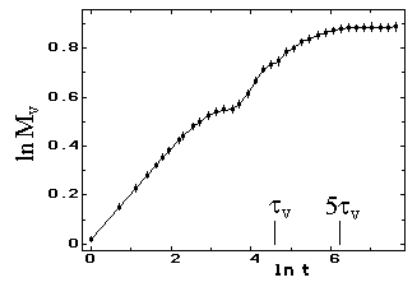
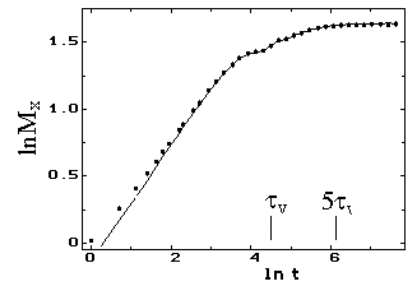
\end{document}